\documentclass[10pt,manuscript,nonacm]{acmart}

\usepackage{longtable}
\usepackage{colortbl}
\usepackage{array}
\usepackage{tabularx}
\usepackage{makecell}
\usepackage{threeparttable}
\usepackage{ragged2e}

\usepackage{mdframed}
\usepackage{xurl}
\usepackage{tikz}
\usetikzlibrary{shapes.geometric,arrows.meta,positioning,calc,fit,backgrounds}

\usepackage[capitalise]{cleveref}

\newcolumntype{Y}{>{\RaggedRight\arraybackslash}X}
\newcolumntype{L}{>{\RaggedRight\arraybackslash}X}
\newcolumntype{C}{>{\centering\arraybackslash}X}

\begin{document}

\title[Am I Just Dumb?]{%
  ``Am I Just Dumb?'': Applicability, Action and Verification in Consumer IoT Security Advice%
}

\author{Veerle van Harten}

\authornote{Corresponding author.}

\orcid{0000-0003-0451-4052}

\affiliation{%
  \department{Technology, Policy, and Management}
  \institution{Delft University of Technology}
  \city{Delft}
  \country{The Netherlands}
}

\email{v.t.c.vanharten@tudelft.nl}

\author{Carlos Hern\'andez Ga\~n\'an}

\affiliation{%
  \department{Technology, Policy, and Management}
  \institution{Delft University of Technology}
  \city{Delft}
  \country{The Netherlands}
}

\author{Michel van Eeten}

\affiliation{%
  \department{Technology, Policy, and Management}
  \institution{Delft University of Technology}
  \city{Delft}
  \country{The Netherlands}
}

\author{Simon Parkin}

\orcid{0000-0002-6667-0440}

\affiliation{%
  \department{Technology, Policy, and Management}
  \institution{Delft University of Technology}
  \city{Delft}
  \country{The Netherlands}
}

\renewcommand{\shortauthors}{van Harten et al.}

\begin{abstract}
  Public campaigns urge people to update their Internet of Things (IoT) devices and change default passwords. What happens when people try? We gave 28 participants in the Netherlands two pieces of government-issued advice and asked them to try applying each to three of six bestselling IoT devices (168 sessions). We located no manufacturer-set password shared across units, the kind the advice describes; the only device-level credential located was unique to its unit. Fewer than half the update sessions established firmware status. Told that a setting might not apply, no participant concluded it did not: they treated whatever related setting the interface offered as the target, and located the difficulty in themselves rather than in the advice or device. Generic advice asks people to judge what only manufacturers can state and only devices can report. Campaigns must be coordinated with device design, or replaced by secure defaults that remove the task.
  
\end{abstract}


\maketitle

\section{Introduction}
\label{sec:introduction}
The widespread adoption of consumer Internet of Things (IoT) devices, including smart cameras and network-connected kitchen appliances, has created new security challenges in the home, as attackers can exploit insecure configurations or vulnerabilities \cite{haney2021s,zeng2017end}. In response, governmental bodies and industry organizations have issued advice on what users should do to secure their devices \cite{barrera2023security,ruthfirst,van2020if,van2025all,blythe2019security}. Two seemingly straightforward messages recur across these campaigns and elsewhere: keep devices up to date and change default passwords \cite{haney2023user,van2025all}. 

Whether people can actually follow such advice remains under-examined. Analysis of national campaigns and manufacturer support materials shows that these recommended features are often not mentioned in device documentation \cite{van2025easier}. Our work joins a body of research that challenges the focus on getting users to \textit{just follow advice} and instead asks how advice can be made reliable \cite{van2025all,herley2009so,redmiles2020comprehensive,stewart2012death,reeves2023generic,reeder2017152}. More generally, Redmiles et al. \cite{redmiles2020comprehensive} examined \textit{perceived} actionability by asking users how difficult they \textit{think} security and privacy advice would be to follow, and Lawo et al. \cite{lawo2026who} surveyed 348 participants on the perceived usability of 29 pieces of IT-security advice and found that variation in these assessments was more closely associated with user characteristics than with text metrics. These studies assess advice as text through readers’ judgments, but do not observe whether the advice works -- i.e., what happens when someone attempts to apply it to a real device. 

We target this gap by conducting a laboratory study of users attempting to apply the advice to popular devices. We first consolidated advice from current national campaigns~\cite{Germany,Australia,Spain,Switzerland,Japan,VeiligInternetten2,US} to create the two advice items: to check whether a device has the most recent updates installed and to change its default password. 
This advice is targeted at IoT (or `smart devices'), which contains a very wide range of devices. It reaches users regardless of the devices they own. To select devices that users would plausibly apply the advice to, we sampled six devices from the Amazon lists of most popular devices in six device categories. Please note that this approach means we did not -- and should not -- pre-screen devices for confirmed feature availability. Our goal is to create a realistic setting that reflects the conditions under which real users interact with the advice.
Participants (n = 28) were assigned three of the six devices and asked to try to apply each of two pieces of advice to every assigned device, resulting in 168 sessions (84 per advice item). Our study data combine observations, think-aloud protocols, interviews, and task workload assessments.
We address two research questions: \textit{(i) What are the outcomes for users attempting to apply common IoT security advice?}, and \textit{(ii) How do device environments and users' pre-existing mental models interact with the advice to shape these outcomes?}

Across the 84 password sessions, 33 reached no password setting, 50 reached an account-level password setting (a credential set on an online account), and one reached a password setting located on the device. In 49 of the 50 sessions that reached an account-level setting, the participant concluded that they had applied the advice; in the remaining one, the participant was unsure. An account password controls access to the vendor account, not the device itself, and is therefore a different credential from the one targeted by the advice. Only one user reached a device-level credential, the closest match to the advice. This password was indeed set by the manufacturer but, in contrast to the default passwords targeted by the advice, it was unique to the individual device. So changing it did not result in the security benefits sought after by the advice.

Two patterns in these outcomes are central to our analysis: participants often had difficulty putting the advice into practice, and reaching a related but different target from the one described by the advice was typically interpreted as having applied it. We examine how these outcomes are related to smart device environment, with non-standardized interfaces, interconnected accounts, and multiple software layers. Because the advice described its target generically, participants had to infer what that target was from interface labels or, when these did not resolve the question, from the support materials. Support was sought in 66 of the 168 sessions. In \Cref{sec:discussion}, we examine how the options exposed by device interfaces shaped outcomes, rather than attributing them primarily to participant characteristics \cite{johnson2012beyond}. We also assessed which target the advice most plausibly referred to.

This study makes three contributions: (i) An account of what generic campaign advice and the devices to which it applies allow people to determine, act on, and verify. (ii) It documents variations in session outcomes across devices and advice items, including cases in which participants’ understanding of what they had done differed from the configuration reached. (iii) It identifies the limits of user-facing advice: device interfaces and support materials can leave users unable to determine whether the advised target is available, whereas some device properties, including per-unit credentials and automatic updating, remove the need for the advised action altogether.

In the following sections, we review related work on IoT security behaviors and advice usability (Section \ref{sec:relwork}). We then detail our methodology (Section \ref{sec:method}), present our results (Section \ref{sec:results}), discuss the implications for advice, devices, and the handover between them (Section \ref{sec:discussion}), and conclude with directions for future work (Section \ref{sec:conc}).

\section{Related Work}
\label{sec:relwork}

\subsection{Security Usability in the Smart Home}
Everyday practices, such as device configuration and maintenance, shape home security outcomes more strongly than stated intentions \cite{haney2021s,turner2021googling,van2025does}. These practices rest with owners, who are not selected for their expertise and are often reluctant to alter settings after installation \cite{haney2021s,nthala2018informal}. Smart device interfaces tend to be minimal \cite{bouwmeester2021thing} and non-standardized \cite{turner2021googling}, and they rarely show the device's security state. Configuration abandonment has also been reported alongside these inconsistencies \cite{turner2021googling}, while rarely exposing either the device's security state or the outcome of a configuration action \cite{haney2020work,bouwmeester2021thing,vetrivel2024iot}. These findings are based on interviews and on owners attempting to remediate known infections, but do not show what a person can determine about a device when trying to apply generic advice to it.
It has been established that interfaces often lack visible status indicators of a smart device's security state \cite{bouwmeester2021thing,whitten1999johnny}. Non-secure behaviors have historically been attributed to weak motivation or low self-efficacy \cite{crossler2014extended,prange2022secure}, although a systematic review of the methods used in the literature found that the link between self-efficacy and behavior is more often assumed than measured \cite{borgert2024self}. Both explanations address whether people are inclined to act, but leave open whether the advised action is available on a particular smart device and whether a person relying on generic advice can determine this.

\subsection{Security Advice and Its Fit with Smart Devices}
Publicly disseminated advice is a common form of formal support for non-experts \cite{reeder2017152,redmiles2020comprehensive}, and campaigns promote actions expected to be the most helpful, such as choosing strong passwords and keeping devices updated \cite{blythe2019security}. Interviews with 21 authors of general security advice describe a production process in which advice is rarely tested with the intended users, and authors tend to overestimate its usability \cite{neil2023comes}. Whether generic advice can be acted upon at all has been questioned on separate grounds \cite{reeves2023generic,stewart2012death}. Smart devices add difficulties of their own. For example, update methods vary from one device to the next \cite{haney2023user}, security features may be left undocumented \cite{van2025easier}, online guidance can conflict \cite{turner2021googling}, and users report uncertainty about their devices' update status \cite{haney2020work}. 

Prior work thus establishes what campaigns recommend, how advice is produced, and what people report doing to secure their devices. None of these studies directly observed what a user could determine about a device, or reach on it, while attempting to apply generic advice.

\section{Methodology}
\label{sec:method}

We observed 28 participants in a laboratory setting in this study. Each participant was asked to try to apply two pieces of consolidated national security advice, changing the default password and checking that the device was up to date, to three of the six consumer devices, resulting in 168 sessions. Participants received the advice as publicly issued, with no caveats or device-specific cues added by us, and interpreted it without navigational or interpretive help from the researchers. We selected devices based on market popularity rather than confirmed feature availability, for the reasons given in \Cref{sec:device_sel}. An overview of the session procedure for each participant is shown in Figure \ref{fig:method_flow}.

\subsection{Experiment Design and Procedure}
\label{sec:experiment_design}
Participants received the two advice items on printed cards (\Cref{sec:advice_sel}; full text in Appendix \Cref{sec:extracted-advice}).

\paragraph{Participant Protocol.} 
\label{sec:procedure}
Participants were asked to try to apply two pieces of security advice (changing the default password and checking that the device is up to date) to three pre-selected devices. At the outset, researchers clarified that some recommended settings might not be applicable to or findable on a given device, and that the objective was to attempt the process. The instructions to change a password or install an update came from the advice itself; the researchers added no further instructions and did not present any sessions as requiring a specific result. Participants were asked to demonstrate the action rather than complete it: they showed how they would set or change a credential without confirming it, and how they would carry out an available update without installing it, so that credentials and firmware versions remained as stable as possible across sessions. The order of the tasks varied across devices and participants. Participants were instructed to imagine that they already owned the devices and were trying to secure them. They were instructed to read and interpret the advice themselves, think aloud while working, use any available materials (smartphones, apps, packaging, manuals, laptops), and complete each task within approximately ten minutes, followed by the National Aeronautics and Space Administration Task Load Index (NASA-TLX) assessment \cite{HartStaveland1988}.

Researchers used neutral, non-leading verbal cues during sessions (for example, “can you tell me what you are thinking?”) to maintain think-aloud verbalization without directing participant behavior, and withheld feedback on task outcomes until post-task debriefing. Clarification requests were deferred to better approximate an unsupported home experience.

\begin{figure}[ht]
\centering
\includegraphics[width=\columnwidth]{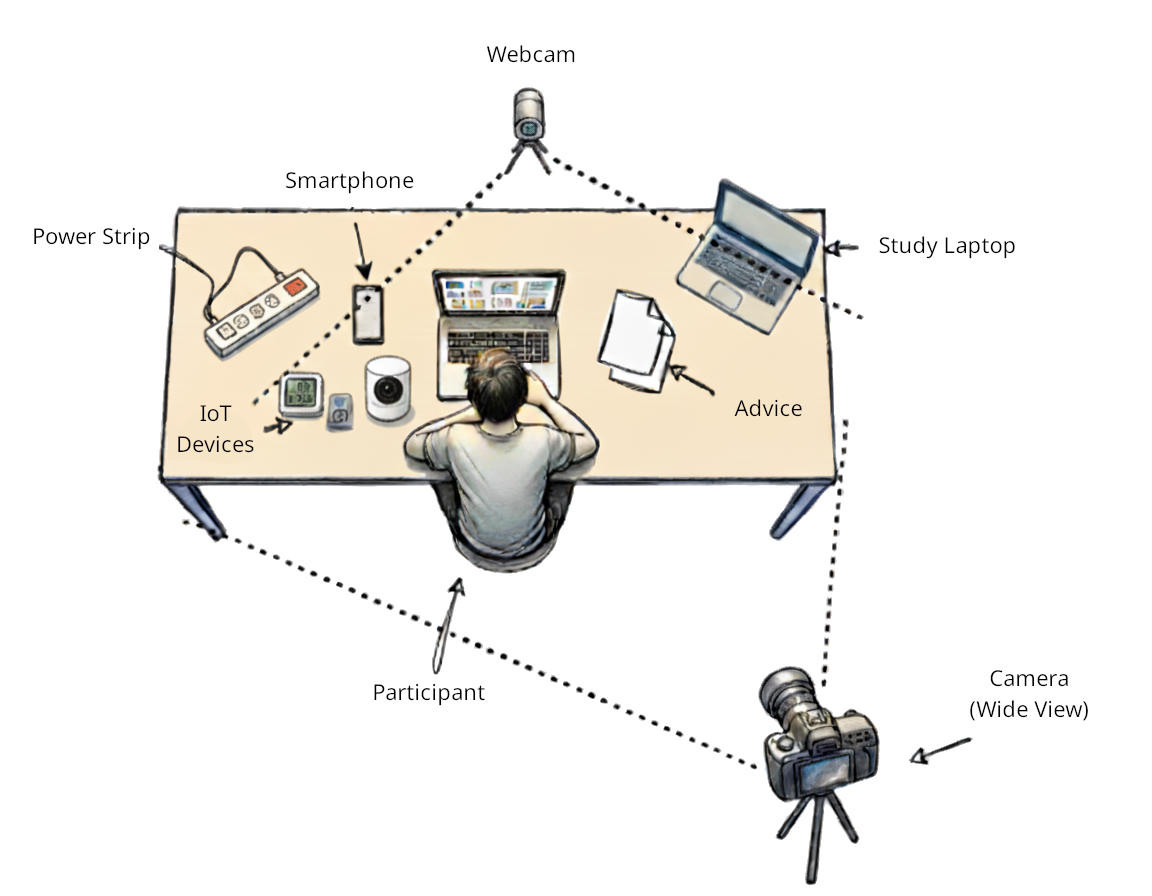}
\captionsetup{font=small}
\caption{Schematic of the study workstation. Participants worked with the assigned IoT devices, a smartphone, and a laptop while reading the two printed advice sheets; two camera angles recorded verbalizations and on-device interactions. The image illustrates the arrangement and was generated using ChatGPT 5, combined with Google’s Nano Banana Image AI, and finetuned in GIMP.}
\Description{Schematic of the study workstation, showing the participant's position relative to the devices, materials, and two cameras. Hand-drawn style, viewed from a three-quarter angle. A participant sits at a desk with two printed advice sheets in front of them, a laptop to the right, and a smartphone to the left. A power strip holding several small IoT devices sits at the far-left edge of the desk. A webcam on a stand points down at the laptop area; a camera on a tripod to the participant's right captures a wide view. Each element is labeled with a leader line.}
\label{fig:setup_diagram}
\end{figure}

We established a ten-minute pragmatic threshold per task. Three pilot sessions with participants from the authors’ network, who had varying levels of technological experience, were conducted to refine the task flow and timing. During these sessions, participants who did not reach the setting within approximately six minutes generally did not reach it at all. To observe participant persistence and ensure that unresolved sessions were not due to a premature cap, the limit was set at ten minutes (600 seconds), above the pilot observation. Participants were permitted to exceed ten minutes on a later task if an earlier session had been completed under the limit, carrying over the unused time. Under this carryover protocol, 33 of the 168 sessions reached or exceeded 600 s, with the longest lasting 973 s. The sessions were recorded using a webcam and tripod-mounted camera to capture verbalizations and interactions with the devices (Figure \ref{fig:setup_diagram}).

\paragraph{In-session assurances.} In seven of the 168 sessions, a researcher, responding to visible distress rather than as an element of the protocol, stated that the devices would be gone through afterward during the debrief (\Cref{sec:posttask}). No navigational or interpretative assistance was provided. 

To keep the sessions focused on the in-use stage of applying advice, all devices were pre-configured (including Wi-Fi connection, study accounts, and app pairing), and this configuration was restored between participants.

\subsection{Post-Task Measures and Debrief}
\label{sec:posttask}
Participants' subjective experiences were captured through the think-aloud protocol, the NASA-TLX scale, and a semi-structured debrief interview. The second part of our analysis related these subjective measures to the outcomes achieved during the sessions.

Immediately following each of the six sessions, the participants completed the raw NASA-TLX scale, retaining all six dimensions but omitting the weighting procedure (Appendix \ref{sec:nasa-tlx}). We used interviews to understand the participants' experiences and reasoning before they were made aware of what we had verified \cite{mckendrick2018deeper}. This allowed us to explore their sense-making, including points of frustration, confidence in their actions, and why they stopped when they did. The full interview protocol is presented in Appendix \ref{fig:debrief_int}.

Every participant was debriefed after the sixth session and the closing interview; all task measures and interviews were completed beforehand. Where a pathway existed, the researcher demonstrated it: the firmware-update pathway on each of the six devices and, for the password advice, the features on the HP DeskJet 2721e and the Tapo C210 that came closest to the description on the card, namely the admin PIN and the optional Camera Account. Where nothing had been located, participants were told in identical terms that the research team had not been able to locate the feature either. Researchers stated that outcomes that did not match the configuration described in the advice stemmed from the relationship between the advice and device design rather than from a lack of skill. 

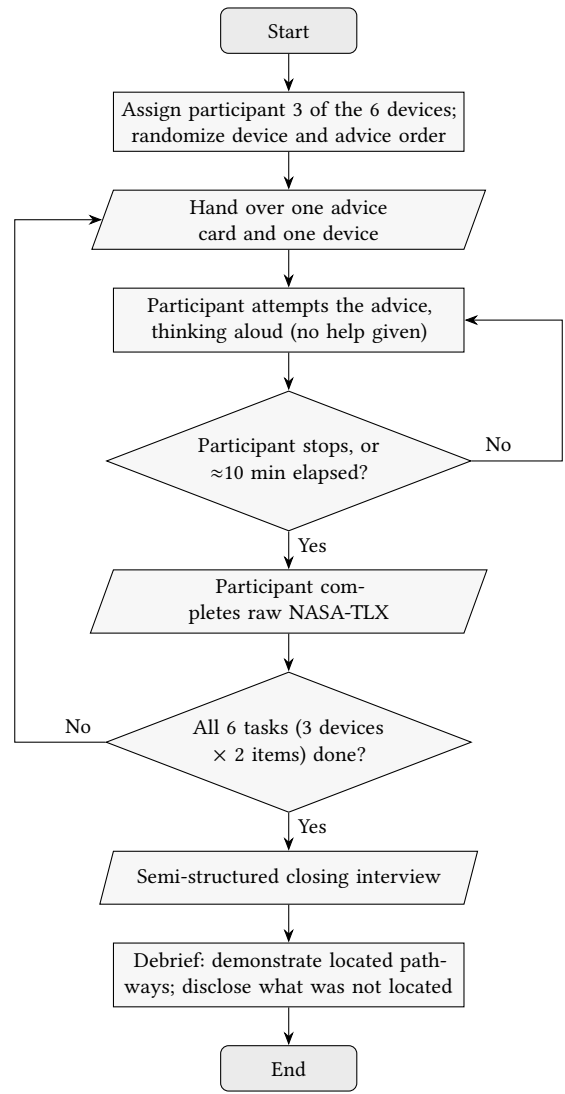
\begin{figure}[t]
\centering
\begin{tikzpicture}[
  font=\footnotesize, node distance=5mm and 10mm,
  term/.style={draw, rounded corners=3pt, fill=black!8, minimum width=18mm, minimum height=6mm, align=center},
  proc/.style={draw, fill=black!3, text width=44mm, minimum height=7mm, align=center},
  io/.style={draw, trapezium, trapezium left angle=70, trapezium right angle=110, trapezium stretches=true, fill=black!3, text width=44mm, minimum height=7mm, align=center},
  dec/.style={draw, diamond, aspect=2.6, fill=black!3, text width=30mm, align=center, inner sep=1pt},
  arr/.style={-{Stealth[length=1.8mm]}, thin}
]
\node[term] (start) {Start};
\node[proc, below=of start] (assign) {Assign participant 3 of the 6 devices; randomize device and advice order};
\node[io, below=of assign] (card) {Hand over one advice card and one device};
\node[proc, below=of card] (attempt) {Participant attempts the advice, thinking aloud (no help given)};
\node[dec, below=of attempt] (stop) {Participant stops, or $\approx$10 min elapsed?};
\node[io, below=of stop] (tlx) {Participant completes raw NASA-TLX};
\node[dec, below=of tlx] (six) {All 6 tasks (3 devices $\times$ 2 items) done?};
\node[io, below=of six] (interview) {Semi-structured closing interview};
\node[proc, below=of interview] (debrief) {Debrief: demonstrate located pathways; disclose what was not located};
\node[term, below=of debrief] (end) {End};
 
\draw[arr] (start) -- (assign);
\draw[arr] (assign) -- (card);
\draw[arr] (card) -- (attempt);
\draw[arr] (attempt) -- (stop);
\draw[arr] (stop) -- node[right, pos=0.35] {Yes} (tlx);
\draw[arr] (stop.east) -- node[above, pos=0.3] {No} ++(12mm,0) |- (attempt.east);
\draw[arr] (tlx) -- (six);
\draw[arr] (six) -- node[right, pos=0.35] {Yes} (interview);
\draw[arr] (six.west) -- node[above, pos=0.3] {No} ++(-12mm,0) |- (card.west);
\draw[arr] (interview) -- (debrief);
\draw[arr] (debrief) -- (end);
\end{tikzpicture}
\captionsetup{font=small}
\caption{Each participant completed six tasks (three devices $\times$ two advice items), rating workload after each task; the interview and debrief took place after the sixth. Devices were preconfigured before each session and restored afterward (\Cref{sec:experiment_design}). The coding procedure for the recorded sessions is described in \Cref{sec:data_analysis}.}
\Description{Flowchart of the session procedure, from assigning devices to the closing debrief, with two loops. Steps in order: (1) assign each participant three of the six devices and randomize device and advice order; (2) hand over one advice card and one device; (3) the participant attempts the advice while thinking aloud, with no help given; (4) a decision point asks whether the participant has stopped or about ten minutes have elapsed; if so, the flow continues to step 5; if not, it returns to step 3; (5) the participant completes the raw NASA-TLX; (6) a decision point asks whether all six tasks are done; if so, the flow continues to step 7; if not, it returns to step 2; (7) a semi-structured closing interview; (8) a debrief in which located pathways are demonstrated and anything not located is disclosed.}
\label{fig:method_flow}
\end{figure}

\subsection{Advice and Device Selection}
\label{sec:advice_sel}
We selected two widely promoted pieces of consumer advice, changing the default password and ensuring devices are up to date, because these are consistently emphasized in governmental and expert advice across many countries \cite{Germany,Australia,Spain,Switzerland,Japan,VeiligInternetten2,US}.

Among these, we focused on the most actionable elements: identifying and changing any pre-configured passwords (excluding strength/reuse considerations) and checking for and applying available firmware updates. The structure of the advice was preserved as it appears in the public domain, reflecting the diverse levels of detail found in current national cybersecurity campaigns: generic guidance on passwords and procedural guidance on updates. Our analysis of the source materials revealed that update advice frequently includes procedural pointers (see Appendix~\ref{sec:traceability-matrix}), whereas password advice typically instructs users to 'change the default' without indicating where to find the relevant settings. We reproduced this asymmetry as issued and compared the two advice items descriptively. The full cards are provided in Appendix~\ref{sec:extracted-advice}.

The password card defines its target as a password that is pre-configured and supplied with the device, is “simple, publicly known, and easily found online,” and is therefore a target for attackers. It instructs the reader to change such a password as soon as possible. The update card provides three steps: open the device app or web interface, find the Settings section, and check the update status. Each piece of advice comprised elements that appeared in at least two national campaigns, with each element traced to its source in Appendix~\ref{sec:traceability-matrix}. 

\subsubsection{Device selection}
\label{sec:device_sel}
The devices were selected to reflect the prevailing consumer trends in the smart home device market. We began with a market analysis of smart home categories and chose Amazon as our sourcing platform because it accounts for a dominant share of European IoT retail, approximately 40\% of the top e-commerce sales \cite{ecdb2025europe}.

On January 13, 2025, we retrieved six devices from the official Amazon ``Best Sellers'' list, which ranks devices by sales volume (see Table~\ref{device_overview_sessions}). The top-ten lists used in our selection did not include any devices marked as ``Sponsored.'' This strategy was chosen to reflect the reality of the consumer market, which often includes a mix of established brands and newer high-volume sellers. For all six devices, selection was based on market popularity rather than on confirmed feature availability. Pre-screening for devices on which the advised action was confirmed to exist would have removed the determination of applicability from the task, which is part of what we set out to observe. Our determination could establish what the available sources reported, but not whether a feature was absent (Section~\ref{sec:prestudy}). Devices had to meet a strict definition of ``smart'': network-connected, remotely controllable, and capable of data exchange. Ambiguous devices that used the word ``smart'' in their product names (e.g., offline smoke alarms) and IoT devices that required additional purchases for basic use (e.g., devices that required a hub) were excluded. Devices were purchased in the Netherlands, with the language set to Dutch where possible, reflecting the national context in which the study was conducted and the local recruitment of participants (see Section~\ref{sec:participant_recruitment}).

Each participant was randomly assigned one of the twenty 3-of-6 device combinations (Table~\ref{device_overview_sessions}). Each combination was used once in the first twenty participants, while P21--P28 were assigned from a reshuffled pool, leaving eight combinations used twice, twelve once, and device totals uneven. Device order was randomized per participant, and advice-item order was independently randomized within each device; neither was counterbalanced. The allocation script is available in the \hyperref[open_science]{Open Science} deposit; it was run unseeded, once per block, and the realized within-session orderings were not recorded.

\begin{table}[ht]
\centering
\footnotesize 
\caption{Selected Devices (84 sessions per advice item; each device was used in twice this number across both items). Categories based on Amazon classification.}
\label{device_overview_sessions}
\begin{tabularx}{\columnwidth}{@{} l X r @{}}
\toprule
\textbf{Category} & \textbf{Device (Model)} & \textbf{Sessions} \\
\midrule
Energy & Smart Plug (Meross) & 16 \\
Household & Hygrometer (SwitchBot) & 14 \\
Comfort \& Lighting & Doorbell (Arlo Chime 2) & 14 \\
Entertainment & Smart Speaker (Amazon Echo Dot) & 14 \\
Networking & Printer (HP Deskjet 2721e) & 13 \\
Security & Indoor Camera (TP-Link Tapo C210) & 13 \\
\bottomrule
\end{tabularx}
\end{table}

\subsubsection{Device determination}
\label{sec:prestudy}
We examined the devices at two points. Before the first session, we determined, for each device and advice item, what a person consulting the interfaces, companion app, manufacturer materials, and public sources could locate (Appendix~\ref{app:sources}). This provided a baseline against which the session endpoints were verified. After data collection, we re-examined the devices and mapped the pathway to each advised target in greater detail, recording the interfaces involved, the number of steps, the navigation depth, and whether the interface used the advice's vocabulary (Appendix~\ref{app:ui_complx}). The second pass describes the pathways participants encountered. 

We report only what these sources returned. Where an undocumented credential or update pathway does not appear, we report what we could not locate, without inferring its absence. 

The first pass produced the following results. A user-accessible firmware update pathway was found in all six devices. On no device did we locate a credential satisfying the advice's definition, that is, one that was pre-configured and supplied with the device, publicly known, and easily found online. Two devices contained a device-level credential facility at the layer addressed by the password advice rather than at the vendor-account level, and these provided the closest match in the sample. On the HP DeskJet 2721e, this was an Embedded Web Server PIN, set by the manufacturer, unique to the individual unit, and printed on the label inside the cartridge access door. The manual identifies it as the password to enter when the EWS prompts for one, and requires it for device-level settings such as the device-level password, as well as for firmware updates \cite{hp2021deskjet}. On the Tapo C210, the feature was an optional Camera Account for local streaming, which was disabled by default, so no credentials were present to change; the camera otherwise paired with the owner's cloud account during the initial configuration. No device-level credential facility was located on the remaining four devices. This determination was not shared with the participants for the reasons given in \Cref{sec:device_sel}.

\subsection{Participants} 
\label{sec:participant_recruitment}
Participants were recruited through a third-party research recruitment agency via a Qualtrics survey that described the study as a paid, in-person opportunity to test popular IoT devices and provide feedback on security advice. The invitation stated that no technical expertise was required and that participants did not need to be current device owners. To assess prior experience, we asked participants during the user sessions whether they had previously used IoT devices at home, at work, or through friends or relatives. 

The sessions took place on the campus of Delft University of Technology, and participants received €100 in store credit, accounting for the recruiter's standard fee and travel costs.

The survey explained that the sessions would last 75–90 min and would include hands-on device use, a post-task survey, and a short interview. The eligibility criteria included being 18 years or older, able to attend in person, and fluent in Dutch or English.

\begin{table}[ht]
\centering
\footnotesize 
\caption{Participant Demographics ($n=28$). In line with Dutch statistics, education was categorized into three groups: low (primary and lower secondary education), middle (intermediate vocational and higher secondary education), and high (higher professional and university education).}
\label{tab:demographics_compact}
\begin{tabularx}{\columnwidth}{@{}l X@{}}
\toprule
\textbf{Category} & \textbf{Distribution (n)} \\
\midrule
\textbf{Gender} & Female (15), Male (13) \\
\textbf{Age} & 24--29 (3), 30--39 (2), 40--49 (6), 50--59 (6), 60--69 (7), 70--75 (4) \\
\textbf{Education} & High (22), Middle (5), Low (1) \\
\textbf{IoT Experience} & Yes (20), No (8) \\
\textbf{Employment} & Full-time (10), Retired (6), Part-time (4), Self-employed (3), Unemployed (2), Other (3) \\
\bottomrule
\end{tabularx}
\end{table}

The final sample (n=28) included individuals with varied employment status and education levels. While education was not used as a proxy for technical skills, the sample was skewed toward higher education: 22 of 28 held higher education, compared to only 6 with middle or low education combined (see Table~\ref{tab:demographics_compact}). Demographic data were supplied by the recruitment provider after the sessions and matched to the participant identifiers. When asked about their IoT device experience, 20 participants reported using at least one IoT device (Table~\ref{tab:demographics_compact}).

\subsection{Ethics}
\label{sec:ethics}
Ethical approval for this study was obtained from the Human Research Ethics Committee (HREC) of Delft University of Technology. The participants provided written informed consent for both the study tasks and the video and audio recordings. Recruitment and participant compensation were managed by a third-party agency, which held the contact details. The research team had access only to randomized participant identifiers and the demographic information reported in Table~\ref{tab:demographics_compact}. The recordings were kept on secure servers at the university and deleted following transcription and de-identification. We recognized that participants might interpret an unresolved session as reflecting a personal shortfall. All participants received the same debriefing on this point (Section~\ref{sec:posttask}), and the quotations included here were stripped of potentially re-identifying context.

\subsection{Data Analysis} 
\label{sec:data_analysis}

Each session produced an annotated transcript and a coded endpoint, both derived from the session itself. Transcripts were annotated against the recordings so that each verbalization was paired with the on-screen action it accompanied, providing a record of the route each session took, distinct from the participant's account. \Cref{fig:method_flow} summarizes the session sequence.

\subsubsection{Session coding}
\label{sec:coding}

Because participants demonstrated rather than carried out the advised action (\Cref{sec:procedure}), each session was assigned a single endpoint: the furthest point reached, in three mutually exclusive categories per advice item. A session reached a setting when the participant arrived at the interface where the advised action would be performed, with only the withheld final step remaining. For the password advice: no password setting reached, an account-level password setting reached, or a device-level password setting. A credential the user chooses for a vendor account lacks the properties by which the advice defines its target (pre-configured and supplied with the device, publicly known, easily found online), so such settings were coded at the account level. For the update advice, the endpoints were: no update reached, a companion app update, or a verified firmware update. A session reached a verified firmware update when the interface reported the device's firmware state, with the version number when available: current, an update available, or updating described as automatic. Because the companion app version did not indicate the device's firmware status, sessions ending at the app were coded separately. Endpoints were read from the on-screen state when the participant stopped, recorded before the next task began, and assigned with reference to the pre-session device determination (\Cref{sec:prestudy}). 

Each session record also included the participant's own conclusion about whether they had applied the advice (\textit{perceived application}: yes, no, or unsure), the duration, the six raw NASA-TLX subscales, and where support materials were consulted, what the source returned, in the same categories as the endpoints. The endpoint records how far the session progressed, independently of these fields; the conclusion records how the participant interpreted that outcome.

\subsubsection{Quantitative analysis}
We report endpoints and workload descriptively, per advice items and device. Following the critique of the NASA-TLX's construct validity and of global scoring \cite{bolton2023mathematical}, we treat the six dimensions as independent ordinal measures reporting Performance (0=\textit{Good}, 100=\textit{Poor} \cite{cao2009nasa}), and Frustration here and the remainder in Appendix~\ref{appendix:tlx_details}. For the update advice, we fitted exploratory mixed-effects logistic models to assess whether a session reached a verified firmware update, with participants as a random intercept. The password endpoint could not be modeled: one of 84 sessions reached a device-level setting. With 28 participants and 13-16 sessions per device, all model results are exploratory (\Cref{sec:limitations}).

\subsubsection{Qualitative analysis}
We analyzed the annotated transcripts and closing interviews using reflexive thematic analysis (RTA) \cite{braun2021one,braun2023doing}, consulting each session's Frustration and Performance ratings as context; the themes do not correspond to TLX scales. The analysis was experiential and primarily inductive. Coding was grounded in participants' language and observed behavior (e.g., \textit{Uncertainty About Existence of Security Feature}, \textit{Trial-and-Error in Troubleshooting}) with existing concepts introduced later to deepen interpretation, a combination Braun and Clarke describe as a continuum rather than a choice \cite{braun2023doing}. During theme development, three concepts served as interpretive lenses rather than hypotheses: advice-device mismatch \cite{bouwmeester2021thing}, acting on a target other than the one the advice intended \cite{whitten1999johnny}, and app/firmware conflation \cite{haney2020work}. Themes were constructed through interpretive synthesis rather than frequency counts (Appendix~\ref{sec:themes}); representative extracts per sub-theme, translated from Dutch, are provided in the supplementary Thematic Evidence Document. 

RTA treats researcher subjectivity as a resource rather than a bias to be contained \cite{braun2021one}. The researcher who ran the sessions translated and coded the transcripts; co-authors did not code independently, and no inter-coder agreement was sought, as RTA does not treat coding agreement as a quality measure \cite{braun2021one}. Team meetings interrogated interpretations rather than sought consensus. The authors work in usable security and security policy research and approached the data disposed to read outcomes as arising from the advice-device interaction rather than from individual skill. We state this position so that the interpretation can be weighed against the extracts and endpoints records on which it is based, which are reported in Appendix~\ref{sec:themes} and the supplementary Thematic Evidence Document.

\section{Results}
\label{sec:results}

This section reports the outcomes of the 168 sessions and participants' judgments of their actions. \Cref{sec:results_overview} reports the password and update sessions in turn, together with participants' own conclusions and their use of support materials. \Cref{sec:results_workload} reports task duration and perceived workload, and \Cref{sec:qualitative} gives a qualitative account of how the sessions unfolded. The interpretation is deferred to \Cref{sec:discussion}.

\begin{table}[t]
\centering
\small
\caption{Session endpoints against participants' own conclusions, by advice item (84 sessions per item). Rows within an item are mutually exclusive; counts are sessions, not participants. Participants demonstrated the advised action without committing a credential or installing firmware (\Cref{sec:procedure}). Per-device counts are given in \Cref{tab:results_overview}.}
\label{tab:endpoint_conclusion}
\begin{tabular}{@{}lcccc@{}}
\toprule
& \multicolumn{3}{c}{\textbf{Participant's own conclusion}} & \\
\cmidrule(lr){2-4}
\textbf{Endpoint reached}
& \textbf{Applied} & \textbf{Unsure} & \textbf{Not applied} & \textbf{Total} \\
\midrule
\multicolumn{5}{@{}l}{\textit{Password advice}} \\
\quad No password setting            &  1 & 0 & 32 & 33 \\
\quad Account-level setting          & 49 & 1 &  0 & 50 \\
\quad Device-level setting\textsuperscript{a} &  1 & 0 &  0 &  1 \\
\cmidrule(l){1-5}
\quad \textbf{Total}                 & \textbf{51} & \textbf{1} & \textbf{32} & \textbf{84} \\
\addlinespace[0.3ex]
\multicolumn{5}{@{}l}{\quad\footnotesize\textit{Memo: matching the card's definition: \textbf{0 of 84}.}} \\
\addlinespace
\multicolumn{5}{@{}l}{\textit{Update advice}} \\
\quad No update                      &  0 & 2 & 25 & 27 \\
\quad Companion-app update           & 19 & 0 &  0 & 19 \\
\quad Verified firmware update       & 35 & 3 &  0 & 38 \\
\cmidrule(l){1-5}
\quad \textbf{Total}                 & \textbf{54} & \textbf{5} & \textbf{25} & \textbf{84} \\
\bottomrule
\end{tabular}

\vspace{0.3em}
{\footnotesize
\raggedright
\textsuperscript{a}The Embedded Web Server PIN on the HP DeskJet 2721e: set by the manufacturer, but unique to the individual unit rather than shared across units (\Cref{sec:prestudy}).
\par}
\end{table}
\begin{table*}[t]
\centering
\small

\caption{Session endpoints and mean durations across 168 sessions (28 participants, six sessions each). Endpoints record the furthest point reached in each session; participants demonstrated the action rather than committing it (Section~\ref{sec:coding}). Unequal device totals reflect the randomized allocation scheme (Section~\ref{sec:device_sel}).}
\label{tab:results_overview}

\begin{tabularx}{\textwidth}{@{}l >{\raggedright\arraybackslash}X cccc@{}}
\multicolumn{6}{@{}l}{\textbf{Panel A. Password Advice}} \\[0.5ex]
\toprule
& & \multicolumn{3}{c}{\textbf{Sessions reaching}} & \\
\cmidrule(lr){3-5}
\textbf{Device}
& \textbf{Credential Type}\textsuperscript{a}
& \makecell{\textbf{No password}\\\textbf{setting}}
& \makecell{\textbf{Account-level}\\\textbf{setting}}
& \makecell{\textbf{Device-level}\\\textbf{setting}}
& \makecell{\textbf{Mean Duration}\\\textbf{(mm:ss)}} \\
\midrule

Meross Smart Plug
& None located
& 4/16
& 12/16
& 0/16
& 4:43 \\

SwitchBot Hygrometer
& None located
& 1/14
& 13/14
& 0/14
& 2:46 \\

Arlo Chime 2
& None located
& 2/14
& 12/14
& 0/14
& 4:01 \\

Amazon Echo Dot
& None located
& 13/14
& 1/14
& 0/14
& 9:57 \\

HP DeskJet 2721e
& Manufacturer-set, per-unit-unique (EWS PIN)
& 9/13
& 3/13
& 1/13
& 7:49 \\

TP-Link Tapo C210
& User-created device account, disabled by default\textsuperscript{b}
& 4/13
& 9/13
& 0/13
& 6:10 \\
\midrule

\textbf{All devices}
& ---
& \textbf{33/84}
& \textbf{50/84}
& \textbf{1/84}
& \textbf{5:51} \\
\bottomrule
\end{tabularx}

\vspace{0.2em}
{\footnotesize
\raggedright
\textsuperscript{a}``None located'' means the pre-study determination identified no such facility in the sources listed in Section~\ref{sec:prestudy}; it does not establish that none exists.\\
\textsuperscript{b}Enabling the local Camera Account requires creating credentials from scratch; no pre-existing credential was present on the device to be modified.
\par}

\vspace{1.0em}

\begin{tabularx}{\textwidth}{@{} >{\raggedright\arraybackslash}X cccc@{}}
\multicolumn{5}{@{}l}{\textbf{Panel B. Update Advice}} \\[0.5ex]
\toprule
& \multicolumn{3}{c}{\textbf{Sessions reaching}} & \\
\cmidrule(lr){2-4}
\textbf{Device}
& \makecell{\textbf{No update}\\\textbf{reached}}
& \makecell{\textbf{Companion-app}\\\textbf{update}}
& \makecell{\textbf{Verified firmware}\\\textbf{update}\textsuperscript{c}}
& \makecell{\textbf{Mean Duration}\\\textbf{(mm:ss)}} \\
\midrule

Meross Smart Plug
& 0/16
& 7/16
& 9/16
& 3:15 \\

SwitchBot Hygrometer
& 7/14
& 3/14
& 4/14
& 6:23 \\

Arlo Chime 2
& 6/14
& 2/14
& 6/14
& 5:22 \\

Amazon Echo Dot
& 7/14
& 2/14
& 5/14
& 7:53 \\

HP DeskJet 2721e
& 6/13
& 4/13
& 3/13
& 6:35 \\

TP-Link Tapo C210
& 1/13
& 1/13
& 11/13
& 3:45 \\
\midrule

\textbf{All devices}
& \textbf{27/84}
& \textbf{19/84}
& \textbf{38/84}
& \textbf{5:29} \\
\bottomrule
\end{tabularx}

\vspace{0.2em}
{\footnotesize
\raggedright
\textsuperscript{c}In two sessions, the companion app initiated a firmware update without the participant selecting one.
\par}

\end{table*}

\subsection{What the Sessions Reached}
\label{sec:results_overview}

\subsubsection{Password sessions}
No session reached a credential matching the class defined by the password advice: pre-configured and supplied with the product, publicly known, and easily found online. Fifty of the 84 password sessions reached a password setting on the vendor account, an owner-created credential that governs the account rather than the device. Thirty-three reached no password setting. One reached a device-level password setting in the Embedded Web Server (EWS) of the HP DeskJet 2721e, accessible only after entering the manufacturer-set PIN printed on the label inside the cartridge access door. That PIN is unique to the individual unit rather than shared across units (\Cref{sec:prestudy}). The 33 sessions that reached no setting spanned 20 participants, whereas the 50 that reached an account-level setting spanned 26; 26 of the 28 participants ended at least one password session at an account credential level.

Table~\ref{tab:results_overview} reports the counts by device. Outcomes clustered at opposite ends: on the Amazon Echo Dot, 13 of 14 sessions reached no password setting, whereas on the SwitchBot Hygrometer, 13 of 14 reached an account-level setting. The HP Deskjet's 13 sessions divide into nine reaching no setting, three reaching an account-level setting, and one the device-level setting.

\subsubsection{Update sessions}
\label{upd_sess}
Of the 84 update sessions, 38 reached a verified firmware update, that is, a session after which inspection established the device's own firmware status; 19 reached an update to the companion app, which does not report firmware status; and 27 reached no update. These three outcomes spanned 21, 12, and 17 participants, respectively.

Counts by device are presented in Table~\ref{tab:results_overview}. Verified firmware updates ranged from 11 of 13 sessions on the Tapo C210 to three of 13 sessions on the HP Deskjet 2721e. On the Meross Smart Plug, the firmware entry and the companion-app update entry sat directly beneath one another in the same menu; all 16 of its sessions reached one or the other (nine firmware and seven companion-app), and none ended with no update.

In six of the 84 update sessions, across three devices, updating proceeded without the participant selecting it: in two, the app began a firmware update on its own (P02, SwitchBot, when opening the firmware menu; P22, Tapo, through a launch prompt); in four, the interface stated that updating happens automatically (P03, P08, P22 on the HP DeskJet's ``Auto-updates: Required''; P01 on the Tapo). All six sessions reached a verified firmware update; the participants concluded that the advice was applied and reported no uncertainty about the resulting state.

\subsubsection{Perceived application}
\label{sec:perceived_app_text}

Participants concluded that they had applied the advice in 51 of the 84 password sessions and 54 of the 84 update sessions. Set against what the sessions reached, these conclusions diverge, and in one direction only (\Cref{tab:endpoint_conclusion}).

In these 51 password sessions, participants had reached an account-level setting (49), the HP DeskJet's per-unit PIN (1) or no password setting at all (1). None had reached the credentials described in the advice. P14 arrived at a credential page rendered non-interactive because authentication ran through a linked Google account; concluded that the change could be made through that account, and rated their performance as good. Of the 54 update sessions, 35 had reached a verified firmware update, and 19 had reached the companion app only.

The divergence did not run the other way. All 57 sessions in which the participant concluded that the advice had not been applied had reached no password setting or no update. Six further sessions ended with the participant unable to say what state the device had reached. One password session (P02, Arlo Chime 2) and five update sessions, three of which had reached a verified firmware update. Arriving at the screen that displays the device's status was not sufficient to be able to report it. 

Participants did question the advice as they worked, objecting that its vocabulary did not match the labels on screen. What none of them questioned was its premise: across all 168 sessions, no participant concluded that the target of the advice might not be present on the device in front of them. Participants who stopped without reaching a setting treated the task as unfinished rather than as inapplicable. The nearest approach was P28, who asked the Echo Dot to change its password and received \emph{``It doesn't support that''}, the assistant's stock reply to an unparsed request rather than a statement about the device. P28 did not treat it as an answer to the question of whether the advice applied. The session continued and ended with no credential setting reached, at 852~s, among the longest in the study. 

\subsubsection{Support materials}
In-app help, manufacturer documentation, and web searches were consulted in 66 of the 168 sessions (40 password sessions and 26 update sessions). Of the 40 password consultations, 21 returned no path, 15 returned a path to an account-level password setting, one returned a path to a device-level password setting, and in three cases, the materials were unavailable. Of the 26 update consultations, 12 returned no path, five returned a path to a companion app update, and nine returned a path to a firmware update. Half of all consultations therefore returned no path, and a further 20 returned a path at a layer other than the one described in the advice. Where support returned the firmware path, all nine update sessions reached a verified firmware update; the single password session in which support returned a device-level path was P22's HP Deskjet session. Consultation also differed within devices: on the Tapo C210, support was consulted in seven of 13 password sessions, five of which returned a path to an account-level setting -- four of those sessions ended there, one ended without reaching any setting -- and in one of 13 update sessions (Section~\ref{sec:what_varied}).

\subsection{Duration and Perceived Workload}
\label{sec:results_workload}
This subsection reports task duration alongside the Performance and Frustration dimensions of the NASA-TLX, which captures subjective task evaluation and affective cost. Following Bolton et al.~\cite{bolton2023mathematical}, we treat the subscales as independent ordinal measures rather than as a composite score; means and medians across all six subscales are provided in Appendix~\ref{appendix:tlx_details}.

\subsubsection{Duration}
\label{sec:results_dur}
Mean session duration varied by device and advice item (\Cref{tab:results_overview}). For the password advice, it ran from 2:46 on the SwitchBot Hygrometer to 9:57 on the Amazon Echo Dot (overall mean 5:51); for the update advice, from 3:15 on the Meross Smart Plug to 7:53 on the Echo Dot (overall mean 5:29). The two orderings differed: the Tapo C210 had the third-longest mean duration for password advice (6:10) and the second-shortest for update advice (3:45). Thirty-three of the 168 sessions ran for 600~s or beyond, the longest at 973~s; the ten-minute limit was administered with the carryover described in \Cref{sec:procedure}.

Duration also varied across endpoints. Password sessions ending with no setting averaged 9:47 (587~s), against 3:14 (194~s) for those ending at an account-level setting, in the same direction on all six devices. Fourteen of the 33 no-setting sessions ran to the ten-minute limit or beyond, against three of the 50 account-level sessions. The single session reaching a device-level setting took 7:20, which was longer than the account-level sessions on that device. At the other end, none of the six sessions in which the device began the update itself or reported updating as automatic (\cref{upd_sess}) approached the ten-minute threshold; the longest ran 4:24 (264 s), against an overall update mean of 5:29.

\subsubsection{Workload subscales}
The highest mean Frustration score occurred for the password advice on the Echo Dot (82.1), the device on which 13 of 14 password sessions reached no password setting (Appendix~\ref{appendix:tlx_details}). The HP Deskjet showed a mean Performance score of 58.1 for password advice (where 0 is good and 100 is poor). Across the 50 password sessions reaching an account-level setting, 39 received a Performance rating of 0, the best available, and the mean was 3.6: participants who reached a credential at the account layer rated their own performance as essentially ideal. The NASA-TLX was administered after each session, so these ratings record how participants evaluated the interaction they had just concluded.

\subsubsection{Exploratory models}
\label{sec:correlation_analysis}

We fitted generalized linear mixed-effects models (GLMMs) to the update endpoint, with participant as a random intercept; both are reported in Appendix~\ref{app:exploratory-models}. Higher Frustration and poorer self-rated Performance are each associated with lower odds of reaching a verified firmware update, but this association rests on the contrast with the 27 sessions that reached no update: within the 57 sessions that reached either update target, neither is distinguishable from zero. Device coefficients are not interpretable at 13 to 16 sessions per device, and the password endpoint could not be modeled, as a single session reached a device-level setting.

\subsubsection{Participant characteristics}
\label{sec:demographics}

Outcome counts by education, age, and prior IoT experience are reported in Appendix~\ref{appendix:demographics}. The subgroups are imbalanced, and each participant contributed six sessions, so subgroup denominators expressed in sessions overstate the number of independent observations.

\subsection{How Sessions Unfolded: Qualitative Analysis}
\label{sec:qualitative}
The thematic analysis of think-aloud protocols and interviews (codebook in Appendix~\ref{sec:themes}) describes what participants were doing along the way. We do not report theme frequencies or their co-occurrence with outcomes. Where we state how many participants made a particular kind of statement, that number describes the sample, not the weight of the theme.

\subsubsection{Determining candidate targets across interface layers}
\label{sec:ambiguous_scope}

The first question participants faced was which architectural layer the advice referred to: the physical device, the vendor account, the companion app, or a linked account with a separate provider such as Google. P02 captured the central ambiguity: \emph{``Is it about the device itself? Or is it about the app?\dots\ I don't know if such a device has its own password or if it's just the same as the app.''} Nothing in the device, app, or support materials resolved this ambiguity. For four of the six devices, most password sessions ended at the account-level setting (Section~\ref{sec:results_overview}). P24 illustrates this: \emph{``Then I see `Name,' `Email login,' `Settings,' so I click on that. Then I see `Change Password.' And I think that means I've completed the task.''}

The update advice posed the same question between the companion app and device firmware, and here the advice's vocabulary and the interfaces diverged. The advice states \emph{check the update status} and \emph{select `Update'}; the interfaces label the \emph{firmware}, a word the update advice never uses. P04: \emph{```Firmware.' What is firmware? That's not software.''} P07: \emph{``I have no idea what firmware is.''} P22 declined to act on it: \emph{``When I see the word `firmware,' I think: oof, that goes really deep into the system. Maybe I shouldn't touch that. What if I break something?''} Uncertainty persisted even after a status screen had been reached. P02, on the Arlo Chime 2: \emph{``I'm a bit uncertain. I don't have 100\% confidence that it's correct because it says `firmware,' and I'm like, is that where I should be looking? Was that the right thing?''} Some revised their reading in the closing interview. P25: \emph{``I had to update the printer, but [\dots] I only updated the app. Not the actual device. So I got that wrong. Same with the other device, by the way.''}

Participants also brought expectations about how updating works. The advice asks the reader to go and check; participants described updating as something that reaches them. P09: \emph{``Updating via the App Store is actually really simple\dots\ But I never thought of it, because everything on my devices happens automatically.''} P10, on the Echo Dot: \emph{``I use an Android phone myself, and then I just go to the Play Store\dots\ Then I'd just update apps, right? And it would show up if needed.''} Only the SwitchBot Hygrometer offered a Dutch interface; sessions were otherwise conducted in Dutch against English-language interfaces (\Cref{sec:procedure}). P21 described the added step: \emph{``Look, I'm really good at English, but I don't read technical English every day. I do understand it, but I have to first put it in the right order in my head, and only then can I apply it.''}

\subsubsection{Locating a setting without signposting}
\label{sec:interfaces}

Across the six devices in these sessions, no interface offered a way to establish that a setting was not present, and sessions that did not locate one ran on rather than stopping. Participants described being stuck in loops: P06 referred to \emph{``circular reasoning''} and P14 to a \emph{``settings loop.''} When the advice's vocabulary and the device's labels did not correspond, navigation became trial and error. P13, on the Echo Dot, searched without a criterion for stopping: \emph{``I'm really checking everything now because I'm not sure if the password is part of this account or related to the Echo Dot itself.''} The update advice does not specify where users should look. Its first step points to either the device app or the configuration menu/web interface, and the final step adds the supplier's website. Interface conventions did not necessarily clarify this when they differed from what participants recognized; a small person icon that did not appear clickable prolonged P02's search on the Arlo Chime 2, and P18 found the account password settings \emph{``under some random thing with no label.''} These problems are not unique to IoT, as similar locate-and-complete breakdowns have been documented in security ceremonies for encrypted chat apps~\cite{vaziripour2017you}.

One session reached a credential stored on the device rather than on the vendor account. It was not the credential the advice described, but it is the closest approximation in the data, and it shows what locating one required. P22 reached the HP DeskJet's EWS through the HP Smart app's ``Advanced Settings,'' which redirected to a web browser, combined with on-screen guidance pointing to the cartridge compartment for a PIN. The route did not carry over: P22's other password sessions ended at an account-level setting, and participants had no way to know beforehand which devices would provide such a route. P05 identified the missing condition directly: \emph{``a universal interface, so that everything works the same way.''}

\subsubsection{Support materials in context}
\label{sec:support_scaffolding}

Participants turned to secondary support, such as help menus, FAQs, and app-specific chatbots, only after in-app exploration and settings had not revealed a pathway. A channel that returned a procedure allowed participants to infer that the device had the feature, regardless of whether the procedure applied to the device in front of them. The Amazon support pages found by several participants described how to update an Echo Dot, whereas P11 followed a video that described a different device. No channel supported the reverse inference, as none stated that a device had no default password or that a feature named in the advice was absent. Half of all consultations returned no path at all, and a further 20 returned a path at a different layer from that named in the advice (\Cref{sec:results_overview}); a participant who found nothing had no way to determine whether the feature was absent or whether the search had simply failed to reach it. Help menus and FAQs addressed initial setup or promotional content. P12 noted: \emph{``I see all kinds of stuff here that I'm not even going to list. ... But nothing related to what you're actually looking for.''} Virtual assistants often misinterpreted queries and looped participants back to the same menus, and printed manuals were described as inaccessible; P27: \emph{``I find manuals annoying, because they're often very long.''}

External search sometimes helped, at the cost of filtering and interpreting third-party advice, with P02 noting reliance on \emph{``multiple sources, both from the makers of the app and from people who ran into the same issues.''} Turner et al.~\cite{turner2021googling} similarly observed that searching for online help returned task-specific manufacturer advice mixed with general advice from third parties. Participants also carried expectations from familiar platforms. P06 expected a searchable settings menu of the kind standard in mobile operating systems: \emph{``Usually, I type what I'm searching for, and it pops up immediately.''} This produced disorientation when, as P02 put it, \emph{``the terms here don't really... They don't quite match what I'm seeing here, you know?''}

\subsubsection{Effort and self-attribution}
\label{sec:frustration_effort}

Password sessions reaching no setting ran roughly three times as long as those ending at an account-level setting, and the highest mean Frustration was recorded for password advice on the Echo Dot, where 13 of 14 sessions reached no setting (Section~\ref{sec:results_workload}). Sessions of this kind ended by abandonment rather than by resolution. P15: \emph{``So yeah, I'm giving up. This one's going back to the store.''} P27 put it in terms of time: \emph{``Ten minutes? I don't have time for that.''} 

In half the cohort, 14 of the 28 participants, that ending was absorbed as a judgment about themselves. Some cast it as a general incapacity: P19's \emph{``Am I just dumb?''}, P21's \emph{``you're dealing with a total tech-illiterate here,''} P18's \emph{``I feel really dumb doing this,''} P23's \emph{``I'm just a layperson.''} Others attributed it to a particular habit or background: P22 to going \emph{``too fast sometimes,''} P15 to not having opened the manual, P06 to having grown up with Apple rather than Android. Both readings locate the difficulty in the participant rather than in the advice or the device.

Self-attribution did not track how participants fared elsewhere. The 14 who attributed difficulty to themselves and the 14 who did not reached comparable endpoints. Thirteen in each group reached an account-level password setting at least once, and 11 and 10 respectively a verified firmware update. P22, who attributed the friction to going \emph{``too fast sometimes,''} reached the study's single device-level setting. Where the mismatch became concrete enough to point at, the attribution moved outward. P19, in the same HP session as \emph{``Am I just dumb?''}, also asked of the advice's second step: \emph{```Go to settings.' Why don't they just add a simple settings tab? \dots\ Where can I find settings?''} Sessions that ended at a status string ended cleanly and without self-attribution. P08, on the Meross: \emph{```Firmware update,' and `0 available updates.' \dots\ Yeah, if it's this simple, even I can understand it.''} Where no status was shown, sessions ended without confirmation, even when they had reached the right place. P13: \emph{``It wasn't clear what exactly was installed on my device\dots\ And even then, you're not sure it's correct.''} P01 named what was missing: \emph{``You miss the confirmation. So that causes a lot of frustration.''} \Cref{sec:posttask} records the debrief statement given to every participant on this point.

\section{Discussion}
\label{sec:discussion}

Our design reproduced the situation that generic advice creates for users. Campaigns issue the same text to all owners of IoT devices without specifying which devices it applies to. Accordingly, we selected devices based on market popularity rather than on confirmed feature availability, and did not tell participants whether either action was available on the device they were given (Section~\ref{sec:device_sel}). Determining whether the advice applied remained part of the task, as it does when users encounter the advice in practice.

In answer to RQ1, the two pieces of advice led to markedly different outcomes. The update advice specifies an entry interface and a sequence of steps: 38 of 84 sessions reached a verified firmware update, 19 ended at a companion-app update, and 27 reached no update. The password advice defines a credential but does not say where to find it. Fifty of 84 sessions ended at an account credential, 33 reached no password setting, and one reached a device-level setting. The latter was in the Embedded Web Server of the HP DeskJet 2721e, behind a manufacturer-set PIN unique to that unit and printed inside its cartridge access door (Section~\ref{sec:what_advice_names}). By the advice's own definition, no session reached the credentials it describes. 

One could defend the advice by holding that recipients must judge for themselves whether it applies to their device. Our participants were in a position to make that judgment. At the outset, they were told that a recommended setting might not be applicable or findable on a given device (Section~\ref{sec:procedure}). In none of the 84 password sessions did a participant conclude that the advice did not apply to the device in hand. They did not reject the advice, which they endorsed; instead, they abandoned their search. In half the cohort, that ending was absorbed as a judgment about themselves rather than as a mismatch between the advice and the device (Section~\ref{sec:frustration_effort}). 

\subsection{What the Advice Left the Recipient to Determine}
\label{sec:what_varied}

Applying either piece of advice required four determinations: whether the advice applied, which target it defined, where to act, and what state the device had reached. The pieces of advice supplied these unevenly. The update advice names a generic pathway and two labels; the password advice defines a credential and stops. Neither states whether it applies to a given device, nor how to confirm the result.

\textbf{Whether it applied.} Participants inferred this from the actions available to them. No source used in the determination reported a credential of the advice's class on any device, and none reported that such a credential was absent (Section~\ref{sec:prestudy}). Where documentation described a device-level credential, which occurred on two of six devices, it did not state whether that credential was shared across units or unique to one, the property that determines whether the advice applies. Therefore, three situations occurred in the sample that were indistinguishable from the outside: a manufacturer-set credential existed but was unique to the unit; a device-level account existed but was disabled, so no credential was present to change it; and on four devices, no such feature was found.

\textbf{Which target it defined.} The target defined by the advice was rarely among those presented by the interface. Instead, participants encountered a vendor account credential or a companion app version number. What participants reached depended on whether the interface made the relevant target visible using the advice's own words. The Tapo C210 illustrates this contrast within a single device: 11 of 13 update sessions reached a verified firmware update, whereas none of the 13 password sessions reached a device-level setting. For updates, the app used the card's vocabulary (``Select `Update','') and displayed a status string (``The firmware of all connected devices is up to date'') that provided a stopping rule for the user. For credentials, the interface presented the vendor account, where nine of the 13 password sessions ended; seven sessions consulted the app's Support Center, user guide, or FAQ -- four of them alongside ending at the vendor account -- and none reached a device-level setting. Matching vocabulary alone was not sufficient: the SwitchBot Hygrometer app also matched the card's wording, yet seven of its 14 update sessions ended without an update. The Echo Dot provides the limiting case: a voice interface offers no browsable surface, making inspection unavailable as a strategy; 13 of its 14 password sessions found no settings.

\textbf{Where to act.} Support materials often did not close the gap: a device-specific path was returned in only one of 40 password consultations and nine of 26 update consultations (Section~\ref{sec:results_overview}). What was missing was not simply a shorter route. Verified firmware updates were reached in 85\% of sessions on the Tapo C210 and 23\% on the HP DeskJet, both at four levels of navigation depth, compared with 36\% on the Echo Dot at one level (Table~\ref{tab:results_overview}; Appendix~\ref{app:ui_complx}). Therefore, navigation depth did not correspond to the outcomes.

\textbf{What state the device had reached.} Where the visible option belonged to a different architectural layer, the setting participants reached governed the vendor account rather than the device. For credentials, we identified no interface in the sample that marked this distinction. For updates, several did, labeling the device-level target \emph{firmware}, a word the advice never uses, so the distinction was available to participants but not in the vocabulary they had been given (Section~\ref{sec:ambiguous_scope}). This reproduces the believed-versus-achieved dissociation described by Whitten and Tygar~\cite{whitten1999johnny}, but here, it arises from the layered smart-home architecture rather than from cryptographic complexity. In P14's HP DeskJet session, the participant reached a credential screen that was non-interactive because authentication ran through a linked account, concluded that the advice had been applied, and rated their performance as good. In answer to RQ2, device environments interact with generic advice such that recipients act on the first familiar application-level pathway they encounter and treat visible account-level completion as confirmation that the advice has been applied.

\subsection{What the Advice Names}
\label{sec:what_advice_names}

The password card defines its target as ``a pre-configured password that comes with your device,'' one that is ``simple, publicly known, and easily found online'' (Appendix~\ref{sec:extracted-advice}). A publicly known credential is one shared across production units, and the three properties listed on the card correspond closely to the credential class that the Mirai botnet brute-forced from standard lists during its first infection stage~\cite{antonakakis2017understanding,bouwmeester2021thing}. The campaign texts do not state this lineage; we reconstruct it, and on that reading, the advice fits an earlier device architecture better than the one encountered here. The available prevalence evidence points in the same direction. Among 270 consumer IoT devices, manufacturer documentation was obtainable for 170; of those, 4.7\% were documented as shipping with a default password, a practice most frequent among smart TVs, a category not represented in our sample~\cite{blythe2019security}. This figure reflects what manuals disclose rather than what devices actually ship with, so it bounds documentation rather than existence. Even so, it provides no basis to expect the described credential on a typical consumer device. The one IP camera in the sample illustrates how far the architecture has shifted, since cameras are the device category most often cited in the default-credential literature~\cite{perone2023default}, yet no preset credential was found on the Tapo C210. Its device-level account is disabled by default and receives a password only when the owner creates one (Section~\ref{sec:prestudy}).

The specificity of the definition is what makes the result more than under-specification. Participants were given a testable definition but had no way of testing it. The card asserts that the credential exists but provides no procedure to determine that it does not, so a search that yields nothing has no defined endpoint. The advice provides no stopping rule, and no interface in the sample supplied one for the credential the password advice defines.

Two of the six devices contained a device-level credential facility, but neither was found to use a shared or publicly known default (Section~\ref{sec:prestudy}). ETSI EN 303 645 requires passwords to be unique per device or user-defined (provision 5.1-1)~\cite{ETSI}, a rule given legal force in the United Kingdom through the PSTI regime~\cite{ukpsti}. The Cyber Resilience Act does not address passwords directly, requiring instead that products be placed on the market with a secure-by-default configuration and with protection from unauthorized access through appropriate control mechanisms (Annex I, Part I)~\cite{CRA2024}. These rules apply to newly placed devices, so both architectures will coexist in the installed base for years, and recipients of generic advice have no way of knowing which one their device uses.

\subsection{What Would Have to Change}
\label{sec:guide_action_or_prompt}

Advice-makers face a choice that has remained implicit thus far. If the goal is to guide a specific configuration, the advice must identify the credential layer and provide recipients a way to determine whether their device has one or not. If the goal is to encourage more general engagement with security~\cite{van2025all}, then the effort observed here is part of the cost, and the relevant question is what that effort achieves. Password changes and firmware maintenance are technical defenses, not symbolic actions~\cite{alrawi2021circle}, and effort directed at the wrong layer does not provide them.

The cost is measurable. Herley~\cite{herley2009so} argued that users rationally reject advice when its costs exceed its benefits; our participants were required to attempt each task, so that option was not available to them. Even with materials at hand, no competing demands, ten minutes per task that could be extended by carrying over unused time (Section \ref{sec:procedure}), 33 of the 168 sessions reached or exceeded the time cap, and 27 of those ended without reaching a setting. For those 27 we know that participants spent at least that long without reaching one; the recorded durations therefore provide a lower bound on the effort elicited by the advice rather than an estimate of the time required for completion. Where no setting was located, there was no completion time for these durations to approximate. Password sessions that reached no setting lasted roughly three times as long as those ending at an account credential, and the highest mean Frustration of the twelve device–task cells was recorded for password advice on the Echo Dot (82.1), where 13 of 14 sessions reached no password setting (Section~\ref{sec:results_workload}; Appendix~\ref{appendix:tlx_details}). This cost is an externality of the advice, borne by recipients under conditions considerably more favorable than a household's.

Neither branch is comfortable. The specific-action branch requires campaigns to provide device knowledge they do not possess; the prompt-attention branch accepts a cost that falls most heavily on those least able to resolve the problem, and our participants turned this into a judgment about themselves. Campaign texts and device documentation share the same omission, as neither tells recipients whether the advice applies to their device. Closing this gap requires coordination between advice and devices, from which three responses follow:

\textbf{1. Declare which security features a device has.} Manufacturers cannot list every feature that a device lacks, because such a list is unbounded. The reverse can be listed: for each action the advice recommends, whether the relevant feature is present or absent. Linking it to the provisions of an existing standard, rather than to campaign wording, gives advice-makers a stable vocabulary. Revising the standard would mean revisiting devices already sold, but the Cyber Resilience Act already requires manufacturers to handle vulnerabilities and publish information about fixes throughout the support period (Annex I, Part II)~\cite{CRA2024}. The format also exists: EN 303 645 Annex B provides an implementation conformance statement, a per-provision record of what a device supports that ETSI allows manufacturers to publish~\cite{ETSI}. Nothing requires it to reach the owner, and existing duties concern updates rather than credentials: the support period must be published (5.3-13), and the Act adds installation instructions and an end date (Annex II). Neither requires the owner to be told whether a device has a device-level password or whether it is shared. Published, the record would give advice somewhere to send its recipients.\\

\textbf{2. Remove the need for the action.} Some device properties eliminate the task rather than ease it, and the regulatory direction points the same way: where a password must be unique per device or user-set, there is no shared default to change, and so no specific action left to guide, for devices newly placed on the market (Section~\ref{sec:what_advice_names}). On the Tapo C210, the local account was disabled by default, so no pre-set credential was present to change (Section~\ref{sec:prestudy}). In six of the 84 update sessions across three devices, the device began the update itself or stated that updating was automatic; in each case, the participant concluded that the advice had been applied and reported no uncertainty (Section~\ref{sec:results_overview}). Reaching the firmware status did not guarantee this: three other sessions reached a verified update and ended with the participant unable to state the device's condition. Six sessions are too few to attribute the difference to device-side action. What is removed is the need for the action, not the action from the advice; the recipient still has to learn that nothing is required, which returns to the first response.\\ 

\textbf{3. Name the layer in words that recipients already have.} The layers participants encountered were the device itself, its companion app, the vendor account, and a linked account with a separate provider (Section~\ref{sec:ambiguous_scope}). Participants raised the distinction themselves: P02 asked whether the device had its own password or shared the app's. What they lacked was a way to resolve it, not the concept. Naming alone is not sufficient: the update interfaces did name the device-level target, calling it \emph{firmware}, a word P04 and P07 did not know, and P22 avoided as risky. We read the breakdowns as terminological rather than navigational, since depth did not correspond to outcomes, while label matches and status strings did. Standardizing interfaces would go further, but our sessions bear on vocabulary and feedback rather than layout, and the requirements above impose uniformity on device properties rather than interfaces. Consistent vocabulary for these layers would allow recipients to transfer what they learn from one device to another~\cite{lin2020transferability}, without requiring uniform iconography or menu structures.\\ 

\subsection{Judging the Advice Rather than the Recipient}

\label{sec:judging_advice}

Half the cohort interpreted the outcome as a personal shortfall, and the question that gives this paper its title arose during a session that ended without reaching a password setting (Section~\ref{sec:frustration_effort}). This interpretation is possible because studies of security advice have largely measured whether people comply and treated non-compliance as a property of the person. Where the advised target is absent or unreachable, the same measurement instead records a limitation of the advice as a shortfall of the person.

Two things follow for the design of such studies. What a device supports can be established and reported before participants are asked to act on it, so that a session ending without the advised target is a finding about the device rather than an open question about the participant. Our determination did this (Section~\ref{sec:prestudy}), which allows us to say that no session reached the credential defined by the advice without asserting that such a credential does not exist. A single measure of success would have also concealed this result. Distinguishing whether the target was identified, whether an action was located, whether the resulting state was verified, and what the participant concluded is what makes the divergence visible: participants concluded that the advice had been applied at an owner-created credential that governs the account rather than the device, and no source available to them marked that difference.

\subsection{Limitations} 
\label{sec:limitations}

The findings are limited to a laboratory study of 28 participants from an urban Dutch cohort, most of whom had above-average formal education (22 of 28) and prior IoT experience (20 of 28). Participants worked in Dutch with five English-language interfaces (Sections~\ref{sec:participant_recruitment} and~\ref{sec:device_sel}). Although non-native English proficiency is high in the Netherlands \cite{engl_profi_2026}, technical terms such as ``firmware'' created conceptual uncertainty, which participants attributed to unfamiliarity with computing concepts rather than to language comprehension (Section~\ref{sec:ambiguous_scope}).

The quantitative modeling is exploratory. Password outcomes showed quasi-complete separation, while device-level odds ratios in the update models had wide confidence intervals (Section~\ref{sec:correlation_analysis}). Therefore, the primary findings rely on the observed endpoints and think-aloud data.

The six devices were selected to observe interaction mechanisms across categories rather than to provide a market census (Section~\ref{sec:device_sel}). Where no device-level credential was located, this reflects what could be verified from the physical device, companion app, and documentation, and does not constitute proof of absence (Section~\ref{sec:prestudy}). Navigation depth and step counts were mapped after data collection, from the device interfaces rather than from session records, to characterize pathway structures rather than serving as controlled experimental manipulations (Appendix~\ref{app:ui_complx}).

Participants demonstrated rather than committed changes, maintaining device states as consistently as possible across sessions (Section~\ref{sec:experiment_design}). Therefore, this study evaluated what participants could determine, reach, and verify, rather than whether they would execute changes to completion. Laboratory observation may also have sustained persistence beyond domestic behavior. Such persistence would, if anything, increase the chance of locating a setting and thus cannot explain the 33 password sessions that ended without one.

\section{Conclusion}
\label{sec:conc} 

This laboratory study examined whether generic consumer IoT security advice enables users to determine applicability, identify targets, reach settings, and verify the resulting state across six bestselling smart-home devices ($N = 28$ participants, 168 sessions). Across 84 password sessions, 33 reached no password setting, 50 reached an account-level setting, and one reached a device-level setting (the per-unit PIN on the HP DeskJet 2721e). On no device in the sample did we locate a manufacturer-set credential shared across units of the kind the advice describes; the one device-level credential we did locate was unique to its unit. Across 84 update sessions, 27 reached no update, 19 reached a companion-app update, and 38 reached a verified firmware update. In 49 of the 50 password sessions ending at account credentials and all 19 update sessions ending at companion-app updates, participants concluded that they had applied the advice given.

For the six devices we studied, the implicit assumption of a common credential and update architecture did not hold. When password advice asserts the existence of a default credential without providing diagnostic criteria, users lack actionable stopping rules. In the absence of interface cues on these devices that differentiate physical hardware credentials from companion app logins and cloud accounts, participants navigated familiar application-level pathways, treating visible account-level completion as confirmation that the advice had been applied. Participants could not determine whether the advice applied, and on the one device that held a pre-set device-level credential, 12 of 13 sessions did not reach the setting at which it is replaced. The advice's own precondition therefore went unmet, and no participant could determine that this was so.

Closing this gap requires coordinating advisory design with device architecture. Where devices incorporate secure defaults or automated maintenance, campaigns should help consumers verify that no manual configuration is required, rather than prompting unnecessary searches. Where manual actions remain necessary, future work should develop diagnostic guidance that enables users to determine whether and where an advised setting exists on their specific hardware, supported by standardized terminology across consumer interfaces that explicitly distinguish device-, application-, and cloud-level controls.

\section*{Use of Generative AI}
Grammarly was used for grammar and spelling correction of author-written text. The study workstation schematic (Figure~\ref{fig:setup_diagram}) was generated using ChatGPT 5 and Google's Nano Banana image model and edited in GIMP.

\section*{Open Science}
\label{open_science}
The de-identified session dataset (comprising 168 sessions and including separate columns for the endpoint reached and the participant's own conclusion, session durations, and the six NASA-TLX subscales), together with the analysis script and output log, the allocation script, and the thematic evidence document, is available at \url{https://osf.io/dm6gn/overview?view_only=4285c4fb2f6b41b5974081fc0100faa9}. The outcome columns should be interpreted using the endpoint definitions in Section~\ref{sec:coding} and the demonstrate-not-complete protocol in Section~\ref{sec:procedure}. Video recordings and transcripts are not deposited.

\bibliographystyle{ACM-Reference-Format}
\bibliography{references_update}

@article{reeder2017152,
  title = {152 simple steps to stay safe online: Security advice for
           non-tech-savvy users},
  author = {Reeder, Robert W and Ion, Iulia and Consolvo, Sunny},
  journal = {IEEE Security \& Privacy},
  volume = {15},
  number = {5},
  pages = {55--64},
  year = {2017},
  publisher = {IEEE},
}

@article{blythe2019security,
  title = {What security features and crime prevention advice is communicated in
           consumer IoT device manuals and support pages?},
  author = {Blythe, John M and Sombatruang, Nissy and Johnson, Shane D},
  journal = {Journal of Cybersecurity},
  volume = {5},
  number = {1},
  pages = {tyz005},
  year = {2019},
  publisher = {Oxford University Press},
}

@article{van2025easier,
  author = {van Harten, Veerle and Hern{\'a}ndez Ga{\~n}{\'a}n, Carlos and
          van Eeten, Michel and Parkin, Simon},
  title = {Unfit for purpose? Assessing the applicability of country-level IoT
           security advice},
  journal = {Journal of Cybersecurity},
  volume = {11},
  number = {1},
  pages = {pp. 1--20},
  year = {2025},
}

@article{barrera2023security,
  title = {Security best practices: a critical analysis using IoT as a case
           study},
  author = {Barrera, David and Bellman, Christopher and van Oorschot, Paul},
  journal = {ACM Transactions on Privacy and Security},
  volume = {26},
  number = {2},
  pages = {1--30},
  year = {2023},
  publisher = {ACM New York, NY},
}

@inproceedings{zeng2017end,
  title = {End user security and privacy concerns with smart homes},
  author = {Zeng, Eric and Mare, Shrirang and Roesner, Franziska},
  booktitle = {Thirteenth Symposium on Usable Privacy and Security (SOUPS 2017)},
  year = {2017},
  pages = {65--80},
  publisher = {USENIX Association},
  address = {Berkeley, CA, USA},
  location = {Santa Clara, CA, USA},
}

@inproceedings{haney2021s,
  title = {``It's the Company, the Government, You and I'': User Perceptions of
           Responsibility for Smart Home Privacy and Security},
  author = {Haney, Julie and Acar, Yasemin and Furman, Susanne},
  booktitle = {30th USENIX security symposium (USENIX Security 21)},
  pages = {411--428},
  year = {2021},
  publisher = {USENIX Association},
  address = {Berkeley, CA, USA},
  location = {Virtual Event},
}

@inproceedings{van2025all,
  title = {“All Sorts of Other Reasons to Do It”: Explaining the Persistence of
           Sub-optimal IoT Security Advice},
  author = {van Harten, Veerle and Hern{\'a}ndez Ga{\~n}{\'a}n, Carlos and
          van Eeten, Michel and Parkin, Simon},
  booktitle = {Proceedings of the 2025 CHI Conference on Human Factors in
               Computing Systems},
  pages = {1--19},
  year = {2025},
  month = apr,
  location = {Yokohama, Japan},
  publisher = {Association for Computing Machinery},
  address = {New York, NY, USA},
  numpages = {19},
}

@inproceedings{ruthfirst,
    author = {Kimberly Ruth and Raymond Buernor Obu and Ifeoluwa Shode and Gavin Li and Carrie Gates and Grant Ho and Zakir Durumeric},
    title = {A First Look at Governments' Enterprise Security Guidance},
    booktitle = {34th USENIX Security Symposium (USENIX Security 25)},
    year = {2025},
    isbn = {978-1-939133-52-6},
    address = {Seattle, WA},
    pages = {119--138},
    url = {https://www.usenix.org/conference/usenixsecurity25/presentation/ruth},
    publisher = {USENIX Association},
    month = aug
}

@article{van2020if,
  title = {What (if any) behaviour change techniques do government-led
           cybersecurity awareness campaigns use?},
  author = {van Steen, Tommy and Norris, Emma and Atha, Kirsty and Joinson, Adam
            },
  journal = {Journal of Cybersecurity},
  volume = {6},
  number = {1},
  pages = {tyaa019},
  year = {2020},
  publisher = {Oxford University Press},
}

@inproceedings{bouwmeester2021thing,
  title = {``The Thing Doesn't Have a Name'': Learning from Emergent Real-World Interventions in Smart Home Security},
  author = {Bouwmeester, Brennen and Rodr{\'\i}guez, Elsa and Ga{\~n}{\'a}n,
            Carlos and van Eeten, Michel and Parkin, Simon},
  booktitle = {Seventeenth Symposium on Usable Privacy and Security (SOUPS 2021)
               },
  pages = {493--512},
  year = {2021},
  publisher = {USENIX Association},
  address = {Berkeley, CA, USA},
  location = {Virtual Event},
}

@misc{Germany,
  author = {{Bundesamt für Sicherheit in der Informationstechnik (BSI)}},
  title = {Smart Home},
  howpublished = {\url{
                  https://www.bsi.bund.de/DE/Themen/Verbraucherinnen-und-Verbraucher/Informationen-und-Empfehlungen/Internet-der-Dinge-Smart-leben/Smart-Home/smart-home_node.html
                  }},
  note = {Last accessed: 06-01-2025},
  year = {n.d.},
}

@misc{VeiligInternetten2,
  author = {{EZK}},
  title = {Hoe moet ik mijn Beveiligingscamera (IP camera) updaten?},
  year = {2024},
  howpublished = {\url{
                  https://veiliginternetten.nl/thema/slimme-apparaten-doe-je-updates/hoe-moet-ik-mijn-beveiligingscamer-ip-camera-up/
                  }},
  note = {(Last accessed: 06-01-2025)},
}

@misc{Australia,
  author = {{Australian Cyber Security Centre}},
  title = {Personal Cyber Security: Advanced Steps},
  howpublished = {\url{
                  https://www.cyber.gov.au/protect-yourself/resources-protect-yourself/personal-security-guides/personal-cyber-security-advanced-steps
                  }},
  note = {Last accessed: 06-01-2025},
  year = {n.d.},
}

@misc{Spain,
  author = {{INCIBE - Instituto Nacional de Ciberseguridad}},
  title = {Dispositivos IoT (Internet de las cosas)},
  howpublished = {\url{
                  https://www.incibe.es/ciudadania/tematicas/dispositivos-iot}},
  note = {Last accessed: 05-01-2025},
  year = {n.d.},
}

@misc{Switzerland,
  author = {{National Cyber Security Centre Switzerland}},
  title = {Cyber Tipp: IoT},
  url = {
         https://www.ncsc.admin.ch/ncsc/de/home/aktuell/im-fokus/2023/cybertipp-iot.html
         },
  organization = {NCSC},
  note = {Last accessed: 07-01-2025},
  year = {n.d.},
}

@misc{Japan,
  author = {{Ministry of Internal Affairs and Communications}},
  title = {End User Security Guidelines},
  howpublished = {\url{
                  https://www.soumu.go.jp/main_sosiki/cybersecurity/kokumin/enduser/enduser_security01_13.html
                  }},
  note = {Last accessed: 07-01-2025},
  year = {n.d.},
}

@inproceedings{mckendrick2018deeper,
  title = {A deeper look at the NASA TLX and where it falls short},
  author = {McKendrick, Ryan D and Cherry, Erin},
  booktitle = {Proceedings of the Human Factors and Ergonomics Society Annual
               Meeting},
  volume = {62},
  pages = {44--48},
  year = {2018},
  publisher = {SAGE Publications},
  address = {Los Angeles, CA, USA},
  location = {Philadelphia, PA, USA},
}

@incollection{braun2023doing,
  title = {Doing reflexive thematic analysis},
  author = {Braun, Virginia and Clarke, Victoria and Hayfield, Nikki and Davey,
            Louise and Jenkinson, Elizabeth},
  booktitle = {Supporting research in counselling and psychotherapy: Qualitative, quantitative, and mixed methods research},
  pages = {19--38},
  year = {2023},
  publisher = {Palgrave Macmillan},
  address = {Cham, Switzerland},
}

@article{braun2021one,
  title = {One size fits all? What counts as quality practice in (reflexive)
           thematic analysis?},
  author = {Braun, Virginia and Clarke, Victoria},
  journal = {Qualitative Research in Psychology},
  volume = {18},
  number = {3},
  pages = {328--352},
  year = {2021},
  publisher = {Taylor \& Francis},
}

@misc{ecdb2025europe,
  author = {{ECDB}},
  title = {Amazon Accounts for 40\% of the Top 20 Online Stores’ Sales},
  year = {2025},
  howpublished = {
                  https://ecdb.com/blog/top-online-stores-in-europe-key-players-market-shares/4647
                  },
  note = {Accessed: 13-01-2025},
  organization = {European eCommerce Market},
}

@inproceedings{turner2021googling,
  title = {When googling it doesn’t work: The challenge of finding security
           advice for smart home devices},
  author = {Turner, Sarah and Nurse, Jason and Li, Shujun},
  booktitle = {International Symposium on Human Aspects of Information Security
               and Assurance},
  pages = {115--126},
  year = {2021},
  publisher = {Springer},
  address = {Virtual Event},
}

@article{van2025does,
  title = {Does protection motivation predict self-protective online behaviour?
           Comparing self-reported and actual online behaviour using a
           population-based survey experiment},
  author = {van’t Hoff-de Goede, MS and Leukfeldt, ER and van de Weijer, SGA and
            van der Kleij, R},
  journal = {Computers in Human Behavior Reports},
  volume = {18},
  pages = {100649},
  year = {2025},
  publisher = {Elsevier},
}

@inproceedings{perone2023default,
  title = {Default credentials vulnerability: The case study of exposed IP cams},
  author = {Perone, Stefano and Faramondi, Luca and Setola, Roberto},
  booktitle = {2023 IEEE International Conference on Cyber Security and
               Resilience (CSR)},
  pages = {406--411},
  year = {2023},
  publisher = {IEEE},
  address = {Piscataway, NJ, USA},
  location = {Venice, Italy},
}

@inproceedings{nthala2018informal,
  title = {Informal support networks: an investigation into home data security
           practices},
  author = {Nthala, Norbert and Flechais, Ivan},
  booktitle = {Fourteenth Symposium on Usable Privacy and Security (SOUPS 2018)},
  pages = {63--82},
  year = {2018},
  publisher = {USENIX Association},
  address = {Berkeley, CA, USA},
  location = {Baltimore, MD, USA},
}

@inproceedings{borgert2024self,
  title = {Self-efficacy and security behavior: results from a systematic review
           of research methods},
  author = {Borgert, Nele and Jansen, Luisa and B{\"o}se, Imke and Friedauer,
            Jennifer and Sasse, M Angela and Elson, Malte},
  booktitle = {Proceedings of the 2024 CHI Conference on Human Factors in
               Computing Systems},
  pages = {1--32},
  year = {2024},
  publisher = {Association for Computing Machinery},
  address = {New York, NY, USA},
  location = {Honolulu, HI, USA},
}

@inproceedings{redmiles2020comprehensive,
  title = {A comprehensive quality evaluation of security and privacy advice on
           the web},
  author = {Redmiles, Elissa M and Warford, Noel and Jayanti, Amritha and Koneru
            , Aravind and Kross, Sean and Morales, Miraida and Stevens, Rock and
            Mazurek, Michelle L},
  booktitle = {29th USENIX Security Symposium (USENIX Security 20)},
  pages = {89--108},
  year = {2020},
  publisher = {USENIX Association},
  address = {Berkeley, CA, USA},
  location = {Virtual Event},
}

@inproceedings{herley2009so,
  title = {So long, and no thanks for the externalities: the rational rejection
           of security advice by users},
  author = {Herley, Cormac},
  booktitle = {Proceedings of the 2009 workshop on New security paradigms
               workshop},
  pages = {133--144},
  year = {2009},
  publisher = {Association for Computing Machinery},
  address = {New York, NY, USA},
  location = {Oxford, United Kingdom},
}

@misc{ETSI,
  author = {ETSI},
  title = {CYBER; Cyber Security for Consumer Internet of Things: Baseline
           Requirements},
  year = {2021},
  url = {
         https://www.etsi.org/deliver/etsi_en/303600_303699/303645/02.01.01_60/en_303645v020101p.pdf
         },
}

@inproceedings{haney2020work,
  title = {Work in progress: Towards usable updates for smart home devices},
  author = {Haney, Julie M and Furman, Susanne M},
  booktitle = {International Workshop on Socio-Technical Aspects in Security and
               Trust},
  pages = {107--117},
  year = {2020},
  publisher = {Springer International Publishing},
  address = {Cham, Switzerland},
  location = {Virtual Event},
}

@article{crossler2014extended,
  title = {An extended perspective on individual security behaviors: Protection
           motivation theory and a unified security practices (USP) instrument},
  author = {Crossler, Robert and B{\'e}langer, France},
  journal = {ACM SIGMIS Database: the DATABASE for Advances in Information
             Systems},
  volume = {45},
  number = {4},
  pages = {51--71},
  year = {2014},
  publisher = {ACM New York, NY, USA},
}

@inproceedings{prange2022secure,
  title = {“Secure settings are quick and easy!”--Motivating End-Users to Choose
           Secure Smart Home Configurations},
  author = {Prange, Sarah and Thiem, Niklas and Fr{\"o}hlich, Michael and Alt,
            Florian},
  booktitle = {Proceedings of the 2022 International Conference on Advanced
               Visual Interfaces},
  pages = {1--9},
  year = {2022},
  publisher = {Association for Computing Machinery},
  address = {New York, NY, USA},
  location = {Frascati, Rome, Italy},
}

@inproceedings{neil2023comes,
  title = {Who comes up with this stuff? interviewing authors to understand how
           they produce security advice},
  author = {Neil, Lorenzo and Ramulu, Harshini Sri and Acar, Yasemin and Reaves,
            Bradley},
  booktitle = {Nineteenth Symposium on Usable Privacy and Security (SOUPS 2023)},
  pages = {283--299},
  year = {2023},
  publisher = {USENIX Association},
  address = {Berkeley, CA, USA},
  location = {Anaheim, CA, USA},
}

@article{stewart2012death,
  title = {Death by a thousand facts: Criticising the technocratic approach to
           information security awareness},
  author = {Stewart, Geordie and Lacey, David},
  journal = {Information Management \& Computer Security},
  volume = {20},
  number = {1},
  pages = {29--38},
  year = {2012},
  publisher = {Emerald Group Publishing Limited},
}

@article{reeves2023generic,
  title = {“Generic and unusable” 1: Understanding employee perceptions of
           cybersecurity training and measuring advice fatigue},
  author = {Reeves, Andrew and Calic, Dragana and Delfabbro, Paul},
  journal = {Computers \& Security},
  volume = {128},
  pages = {103137},
  year = {2023},
  publisher = {Elsevier},
}

@inproceedings{haney2023user,
  title = {User perceptions and experiences with smart home updates},
  author = {Haney, Julie M and Furman, Susanne M},
  booktitle = {2023 IEEE Symposium on Security and Privacy (SP)},
  pages = {2867--2884},
  year = {2023},
  publisher = {IEEE},
  address = {Piscataway, NJ, USA},
  location = {San Francisco, CA, USA},
}

@inproceedings{vaziripour2017you,
  title = {Is that you, Alice? A usability study of the authentication ceremony
           of secure messaging applications},
  author = {Vaziripour, Elham and Wu, Justin and O'Neill, Mark and Whitehead,
            Jordan and Heidbrink, Scott and Seamons, Kent and Zappala, Daniel},
  booktitle = {Thirteenth Symposium on Usable Privacy and Security (SOUPS 2017)},
  pages = {29--47},
  year = {2017},
  publisher = {USENIX Association},
  address = {Berkeley, CA, USA},
  location = {Santa Clara, CA, USA},
}

@inproceedings{whitten1999johnny,
  title = {Why Johnny Can't Encrypt: A Usability Evaluation of PGP 5.0},
  author = {Whitten, Alma and Tygar, J Doug},
  booktitle = {USENIX security symposium},
  pages = {169--184},
  year = {1999},
  publisher = {USENIX Association},
  address = {Berkeley, CA, USA},
  location = {Washington, D.C., USA},
}

@inproceedings{alrawi2021circle,
  author = {Omar Alrawi and Charles Lever and Kevin Valakuzhy and Ryan Court and
            Kevin Snow and Fabian Monrose and Manos Antonakakis},
  title = {The circle of life: A large-scale study of the IoT
           malware lifecycle},
  booktitle = {30th USENIX Security Symposium (USENIX Security 21)},
  year = {2021},
  isbn = {978-1-939133-24-3},
  pages = {3505--3522},
  url = {
         https://www.usenix.org/conference/usenixsecurity21/presentation/alrawi-circle
         },
  publisher = {USENIX Association},
  address = {Berkeley, CA, USA},
  location = {Virtual Event},
  month = aug,
}

@article{bolton2023mathematical,
  title = {The mathematical meaninglessness of the NASA task load index: A level
           of measurement analysis},
  author = {Bolton, Matthew L and Biltekoff, Elliot and Humphrey, Laura},
  journal = {IEEE Transactions on Human-Machine Systems},
  volume = {53},
  number = {3},
  pages = {590--599},
  year = {2023},
  publisher = {IEEE},
}

@article{cao2009nasa,
  title = {NASA TLX: Software for assessing subjective mental workload},
  author = {Cao, Alex and Chintamani, Keshav K and Pandya, Abhilash K and Ellis,
            R Darin},
  journal = {Behavior Research Methods},
  volume = {41},
  number = {1},
  pages = {113--117},
  year = {2009},
  publisher = {Springer},
}

@misc{CRA2024,
  author = {{European Parliament and Council of the European Union}},
  title = {Regulation ({EU}) 2024/2847 on horizontal cybersecurity requirements
           for products with digital elements ({Cyber Resilience Act})},
  year = {2024},
  url = {https://eur-lex.europa.eu/eli/reg/2024/2847/oj},
}

@inproceedings{lin2020transferability,
  title = {Transferability of privacy-related behaviours to shared smart home
           assistant devices},
  author = {Lin, Vanessa Z and Parkin, Simon},
  booktitle = {2020 7th International Conference on Internet of Things: Systems,
               Management and Security (IOTSMS)},
  pages = {1--8},
  year = {2020},
  publisher = {IEEE},
  address = {Piscataway, NJ, USA},
  location = {Paris, France},
}

@misc{US,
  author = {CISA},
  year = {2020},
  title = {Securing the Internet of Things (IoT)},
  howpublished = {\url{
                  https://www.cisa.gov/news-events/news/securing-internet-things-iot
                  }},
  note = {Last updated: 17-12-2020},
}

@online{engl_profi_2026,
  author = {{World Population Review}},
  title = {EF English Proficiency Index by Country 2026},
  year = {2026},
  url = {    https://worldpopulationreview.com/country-rankings/ef-english-proficiency-index-by-country
         },
  urldate = {2026-04-03},
  organization = {World Population Review},
}

@article{johnson2012beyond,
  author  = {Johnson, Eric J. and
             Shu, Suzanne B. and
             Dellaert, Benedict G. C. and
             Fox, Craig and
             Goldstein, Daniel G. and
             H{\"a}ubl, Gerald and
             Larrick, Richard P. and
             Payne, John W. and
             Peters, Ellen and
             Schkade, David and
             Wansink, Brian and
             Weber, Elke U.},
  title   = {Beyond nudges: Tools of a choice architecture},
  journal = {Marketing Letters},
  year    = {2012},
  month   = jun,
  volume  = {23},
  number  = {2},
  pages   = {487--504},
  doi     = {10.1007/s11002-012-9186-1},
  url     = {https://doi.org/10.1007/s11002-012-9186-1},
  issn    = {1573-059X}
}

@incollection{HartStaveland1988,
  author    = {Hart, Sandra G. and Staveland, Lowell E.},
  title     = {Development of {NASA-TLX} ({Task Load Index}): Results of Empirical and Theoretical Research},
  booktitle = {Human Mental Workload},
  editor    = {Hancock, Peter A. and Meshkati, Najmedin},
  series    = {Advances in Psychology},
  volume    = {52},
  pages     = {139--183},
  publisher = {North-Holland},
  year      = {1988},
  doi       = {10.1016/S0166-4115(08)62386-9},
  url       = {https://www.sciencedirect.com/science/article/pii/S0166411508623869}
}

@manual{hp2021deskjet,
  author = {{HP Development Company, L.P.}},
  title  = {{HP DeskJet 2700e All-in-One Series User Guide}},
  year   = {2021},
  url    = {https://h10032.www1.hp.com/ctg/Manual/c07043348.pdf}
}

@misc{ukpsti,
  author       = {{Department for Science, Innovation and Technology}},
  title        = {{The UK Product Security and Telecommunications Infrastructure (Product Security) Regime}},
  year         = {2023},
  month        = apr,
  day          = {29},
  howpublished = {\url{https://www.gov.uk/government/publications/the-uk-product-security-and-telecommunications-infrastructure-product-security-regime}},
  note         = {Last updated 2 May 2024}
}

@inproceedings{antonakakis2017understanding,
  author = {Manos Antonakakis and Tim April and Michael Bailey and Matt Bernhard
            and Elie Bursztein and Jaime Cochran and Zakir Durumeric and J. Alex
            Halderman and Luca Invernizzi and Michalis Kallitsis and Deepak
            Kumar and Chaz Lever and Zane Ma and Joshua Mason and Damian
            Menscher and Chad Seaman and Nick Sullivan and Kurt Thomas and Yi
            Zhou},
  title = {Understanding the {Mirai} Botnet},
  booktitle = {26th USENIX Security Symposium (USENIX Security 17)},
  year = {2017},
  isbn = {978-1-931971-40-9},
  pages = {1093--1110},
  publisher = {USENIX Association},
  address = {Vancouver, BC, Canada},
  url = {https://www.usenix.org/conference/usenixsecurity17/technical-sessions/presentation/antonakakis},
}

@inproceedings{lawo2026who,
  author    = {Lawo, Dennis and Hindrichs, Jenny and Stevens, Gunnar},
  title     = {Who Can Actually Follow {IT}-Security Advice? Exploring the Usability of {IT}-Security Advice},
  booktitle = {Proceedings of the Twenty-Second Symposium on Usable Privacy and Security (SOUPS 2026)},
  year      = {2026},
  publisher = {USENIX Association},
  address   = {Hannover, Germany},
  pages     = {685--704},
  url       = {https://www.usenix.org/conference/soups2026/presentation/lawo}
}

@inproceedings{vetrivel2024iot,
author = {Swaathi Vetrivel and Brennen Bouwmeester and Michel van Eeten and Carlos H. Ganan},
title = {{IoT} Market Dynamics: An Analysis of Device Sales, Security and Privacy Signals, and their Interactions},
booktitle = {33rd USENIX Security Symposium (USENIX Security 24)},
year = {2024},
isbn = {978-1-939133-44-1},
address = {Philadelphia, PA},
pages = {7031--7048},
url = {https://www.usenix.org/conference/usenixsecurity24/presentation/vetrivel},
publisher = {USENIX Association},
month = aug
}

\clearpage
\appendix
\onecolumn
\section{Instructions for User Study}
\label{sec:extracted-advice}

\vspace{10pt}

\begin{mdframed}[
  linewidth=1pt,
  roundcorner=10pt,
  backgroundcolor=gray!10,
  innertopmargin=10pt,
  innerbottommargin=10pt
]
    \section*{Card: Default Password}

    A default password is a pre-configured password that comes with your device. These passwords are often simple, publicly known, and easily found online, making them a prime target for hackers. Leaving the default password unchanged can have serious consequences, including unauthorized access to your personal information, the installation of harmful malware, and even the possibility of your device being used in cyberattacks. To protect yourself and your data, it is important to change the default password as soon as possible.
\end{mdframed}
\vspace{10pt}

\begin{mdframed}[
  linewidth=1pt,
  roundcorner=10pt,
  backgroundcolor=gray!10,
  innertopmargin=10pt,
  innerbottommargin=10pt
]
    \section*{Card: Updates}

    Manufacturers frequently release updates to improve the functionality and 
    security of their smart devices. These updates help fix vulnerabilities, 
    protect against cyber threats, and ensure that the device runs smoothly. 
    To check if your device is up-to-date, follow the steps below:

    \begin{itemize}
        \item \textbf{Access your device:} 
              Open the device app or the configuration menu/web interface.
        \item \textbf{Find Settings:} 
              Locate the 'Settings' section.
        \item \textbf{Check the update status:} 
              Select 'Update' or navigate to the updates section on the supplier's website to check if the latest update has been installed. If a new update is available, install it.
    \end{itemize}
\end{mdframed}
\vspace{10pt}

\raggedright
\small 
\textit{Participants in the user study received guidance on applying two pieces of best-practice security advice, derived from official government websites. This consolidated advice, presented on separate cards, offered clear instructions on how to update device software and change default passwords.}

\newpage
\section{NASA-TLX Scale}
\label{sec:nasa-tlx}

\newcommand{\drawscale}[3]{
    \draw (0,0) -- (10,0); 
    \foreach \x in {0,1,...,10} {
        \draw (\x,0.2) -- (\x,-0.2); 
    }
    \node at (0,-0.5) {#1};
    \node at (10,-0.5) {#2};
    \node at (5,-0.75) {#3}; 
}

\begin{center}
    \resizebox{\textwidth}{!}{ 
    \begin{tikzpicture}
    \node at (0,0) {MENTAL DEMAND};
    \begin{scope}[shift={(6,0)}]
        \drawscale{Low}{High}{}
    \end{scope}
    
    \begin{scope}[shift={(0,-2)}]
    \node at (0,0) {PHYSICAL DEMAND};
    \begin{scope}[shift={(6,0)}]
        \drawscale{Low}{High}{}
    \end{scope}
    \end{scope}

    \begin{scope}[shift={(0,-4)}]
    \node at (0,0) {TEMPORAL DEMAND};
    \begin{scope}[shift={(6,0)}]
        \drawscale{Low}{High}{}
    \end{scope}
    \end{scope}

    \begin{scope}[shift={(0,-6)}]
    \node at (0,0) {EFFORT};
    \begin{scope}[shift={(6,0)}]
        \drawscale{Low}{High}{}
    \end{scope}
    \end{scope}

    \begin{scope}[shift={(0,-8)}]
    \node at (0,0) {PERFORMANCE};
    \begin{scope}[shift={(6,0)}]
        \drawscale{Good}{Poor}{}
    \end{scope}
    \end{scope}

    \begin{scope}[shift={(0,-10)}]
    \node at (0,0) {FRUSTRATION};
    \begin{scope}[shift={(6,0)}]
        \drawscale{Low}{High}{}
    \end{scope}
    \end{scope}
    \end{tikzpicture}
    }
\end{center}

\section{Detailed NASA-TLX Subscale Results}
\label{appendix:tlx_details}

Table \ref{tab:appendix_tlx} provides the full breakdown of the NASA-TLX subscale scores for all six devices across both types of security advice. Following the critique by \cite{bolton2023mathematical}, we present these as independent dimensions rather than a composite score.

\begin{table}[ht]
\centering
\scriptsize
\setlength{\tabcolsep}{3pt}
\caption{NASA-TLX Subscale Mean Scores per Device (N=28 participants, 168 sessions).}
\label{tab:appendix_tlx}
\begin{tabularx}{\columnwidth}{@{} l XXXXXX @{}}
\toprule
\textbf{Device} & \textbf{Ment.} & \textbf{Phys.} & \textbf{Temp.} & \textbf{Effort} & \textbf{Perf.*} & \textbf{Frust.} \\
\midrule
\rowcolor[HTML]{F2F2F2} 
\multicolumn{7}{l}{\textbf{Adv-Password (Change Default Password)}} \\
A. Meross Smart Plug  & 40.6 & 39.1 & 44.1 & 42.2 & 26.3 & 35.6 \\
B. SwitchBot Hygrometer & 24.3 & 23.6 & 23.6 & 24.6 & 2.1  & 10.0 \\
C. Arlo Chime 2         & 35.7 & 31.8 & 41.4 & 40.7 & 17.6 & 29.6 \\
D. Amazon Echo Dot      & 72.9 & 76.4 & 95.0 & 85.4 & 89.3 & 82.1 \\
E. HP Deskjet 2721e     & 57.7 & 60.8 & 81.9 & 70.0 & 58.1 & 59.2 \\
F. Tapo C210            & 50.4 & 51.2 & 61.2 & 57.3 & 17.7 & 53.8 \\
\midrule
\rowcolor[HTML]{F2F2F2} 
\multicolumn{7}{l}{\textbf{Adv-Update (Ensure Device is Up-to-Date)}} \\
A. Meross Smart Plug  & 34.7 & 30.3 & 37.5 & 34.1 & 2.2  & 28.1 \\
B. SwitchBot Hygrometer & 61.8 & 60.4 & 59.3 & 61.8 & 60.0 & 57.1 \\
C. Arlo Chime 2         & 46.4 & 53.2 & 60.0 & 56.8 & 48.9 & 52.5 \\
D. Amazon Echo Dot      & 66.4 & 62.5 & 72.1 & 60.4 & 45.0 & 59.3 \\
E. HP Deskjet 2721e     & 57.3 & 47.7 & 64.2 & 60.0 & 51.2 & 55.8 \\
F. Tapo C210            & 28.5 & 17.3 & 28.1 & 23.1 & 10.4 & 23.5 \\
\bottomrule
\end{tabularx}
\smallskip
\textit{Note: All scales are 0--100. Higher scores indicate more negative outcomes throughout: greater demand, effort, and frustration, and poorer performance. The Performance subscale was administered anchored Good (0) to Poor (100), so it already runs in the same direction as the others and required no transformation.}
\end{table}
\newpage
\section{Debrief Interview Questions}
\label{fig:debrief_int}

\subsection*{1. Initial Impressions}
\begin{itemize}
    \item What were your first thoughts upon reading each piece of advice?
\end{itemize}

\subsection*{2. Knowledge and Skills}
\begin{itemize}
    \item Where do you typically learn about how to secure your devices? (e.g., manufacturer instructions, online searches, friends/family, workplace training, social media, other sources?)
\end{itemize}

\subsection*{3. Application Process}
\begin{itemize}
    \item What challenges, if any, did you face when applying the advice?
    \item How did you attempt to overcome these challenges?
\end{itemize}

\subsection*{4. Resource Utilization}
\begin{itemize}
    \item What types of resources do you find most helpful when setting up or securing new tech devices?
\end{itemize}

\subsection*{5. Device Support Materials}
\begin{itemize}
    \item Did the device’s support materials help you apply the advice? Why or why not? What specific improvements would make them more helpful?
\end{itemize}

\subsection*{6. Security Importance and Motivation}
\begin{itemize}
    \item When setting up a device, do you prioritize security over ease of use or cost? Why?
\end{itemize}

\subsection*{7. Overall Experience}
\begin{itemize}
    \item What was the most valuable thing you learned from this experience?
    \item What additional thoughts or suggestions do you have on how users can be better supported in securing their devices?
\end{itemize}

\newpage
\section{Analysis of Information Sources}
\label{app:sources}

\begin{table}[htbp]
\centering
\begin{threeparttable}
\caption{Results of the pre-study determination (\Cref{sec:prestudy}): what a person consulting the device interface and 14 documentary sources per device and advice item could locate before data collection. The documentary sources comprised the manual, the quick-start guide, the manufacturer website, a Google AI-generated overview, the first five non-sponsored Google results, and the first five YouTube results. The password criterion is reported in the two layers used in the determination: the class defined by the advice card (a credential preconfigured and supplied with the product, simple, publicly known, and easily found online) and the broader class of any facility through which a credential authenticates access to the device itself, as distinct from a vendor account. No product in the sample carried a credential satisfying the card's definition. Where a cell reads ``None located'', this records what these sources returned and does not establish that no such credential or pathway exists; the two facilities that were located are reported accordingly. $n$ gives the number of the 14 documentary sources in which the item was found.}
\label{tab:adv_existence}
\small
\begin{tabular}{llclclc}
\toprule
& \multicolumn{2}{c}{Card-defined default} & \multicolumn{2}{c}{Device-level credential} & \multicolumn{2}{c}{Firmware update} \\
& \multicolumn{2}{c}{credential} & \multicolumn{2}{c}{facility} & \multicolumn{2}{c}{pathway} \\
\cmidrule(lr){2-3} \cmidrule(lr){4-5} \cmidrule(lr){6-7}
Device & Determination & $n$ & Determination & $n$ & Determination & $n$ \\
\midrule
Meross Smart Plug     & None located & 0 & None located          & 0 & Located & 4  \\
SwitchBot Hygrometer  & None located & 0 & None located          & 0 & Located & 5  \\
Arlo Chime 2          & None located & 0 & None located          & 0 & Located & 3  \\
Amazon Echo Dot       & None located & 0 & None located          & 0 & Located & 9  \\
HP DeskJet 2721e      & None located & 0 & Located\tnote{a}      & 2 & Located & 2  \\
TP-Link Tapo C210     & None located & 0 & Located\tnote{b}      & 2 & Located & 10 \\
\midrule
All devices & None located & 0 & Located on 2 of 6 & --- & Located on 6 of 6 & --- \\
\bottomrule
\end{tabular}
\begin{tablenotes}[flushleft]\footnotesize
\item[a] The Embedded Web Server PIN: manufacturer-set, unique to the individual unit,
and printed on the label inside the cartridge access door. The mechanism is documented
by the manufacturer; the value is not published
(\Cref{sec:prestudy}).
\item[b] The Camera Account for local streaming: disabled at factory default, so no
credential was present on the device. Enabling it requires the owner to create a
username and password (\Cref{sec:prestudy}).
\item The device interface was examined in addition to the 14 documentary sources and
is not counted in $n$.
\end{tablenotes}
\end{threeparttable}
\end{table}

\newpage
\section{Exploratory Models of the Verified-Firmware Endpoint}
\label{app:exploratory-models}
Both models use the same binary endpoint, whether a session reached a verified firmware update, which occurred in 38 of 84 update sessions. The reference category combines 19 sessions that reached a companion-app update with 27 that reached no update, although these groups differ. Median Frustration was 25 in companion-app sessions and 90 in no-update sessions, compared with 10 in firmware sessions. Median duration was 195\,s and 563\,s, respectively, compared with 131\,s, with 13 of the 27 no-update sessions running to or beyond the ten-minute cap (\Cref{sec:results_dur}). Because companion-app sessions resembled firmware sessions on both measures, the association rests on the contrast with no-update sessions. Among the 57 sessions that reached either update object, neither subscale was distinguishable from zero (Frustration OR $= 0.61$, 95\% CI [0.31, 1.21]; Performance OR $= 0.85$, [0.28, 2.59]). Medians are calculated over sessions rather than participants. Because the NASA-TLX was completed after each session, the coefficients are associational. Password sessions are not modeled (\Cref{sec:correlation_analysis}).
\begin{table}[ht]
\centering
\small
\renewcommand{\arraystretch}{1.1}
\begin{threeparttable}
\caption{Exploratory mixed-effects logistic models of one update-session endpoint, reaching a verified firmware update (38 of 84 sessions). Both models include participant as a random intercept, so the odds ratios are subject-specific. Frustration and Performance are entered in separate models because they are collinear ($r = 0.75$), with device entered in full in each model. The subscales are modeled separately rather than averaged into a composite, following Bolton et al.~\cite{bolton2023mathematical}.}
\label{tab:exploratory-update-models}
\begin{tabularx}{\columnwidth}{@{} X c c r @{}}
\toprule
\textbf{Term} & \textbf{OR} & \textbf{$p$} & \textbf{95\% CI} \\
\midrule
\multicolumn{4}{l}{\textit{Panel A: Frustration} (AIC 88.3)} \\
Intercept (Meross, 9/16) & 0.67 & .712 & [0.08, 5.68] \\
SwitchBot Hygrometer (4/14) & 0.54 & .659 & [0.03, 8.44] \\
Arlo Chime 2 (6/14) & 0.78 & .863 & [0.05, 13.09] \\
Amazon Echo Dot (5/14) & 0.96 & .974 & [0.07, 13.97] \\
HP DeskJet 2721e (3/13) & 0.18 & .303 & [0.01, 4.66] \\
Tapo C210 (11/13) & 27.36 & .101 & [0.52, 1426.8] \\
Frustration\tnote{a} & 0.09 & .016 & [0.01, 0.64] \\
\addlinespace[0.8em]
\multicolumn{4}{l}{\textit{Panel B: Self-rated performance} (AIC 83.9)} \\
Intercept (Meross) & 0.15 & .186 & [0.01, 2.48] \\
SwitchBot Hygrometer & 8.27 & .281 & [0.18, 385.2] \\
Arlo Chime 2 & 2.92 & .536 & [0.10, 87.1] \\
Amazon Echo Dot & 2.83 & .546 & [0.10, 83.2] \\
HP DeskJet 2721e & 0.56 & .729 & [0.02, 14.43] \\
Tapo C210 & 120.88 & .070 & [0.68, 21574.9] \\
Performance\tnote{b} & 0.05 & .008 & [0.005, 0.44] \\
\bottomrule
\end{tabularx}
\begin{tablenotes}[flushleft]
\footnotesize
\item[a] Standardized NASA-TLX subscale, administered after the session; higher values indicate
greater frustration.
\item[b] Standardized NASA-TLX subscale, administered after the session, scored so that higher
values indicate poorer self-rated performance.
\item \textit{Note:} 84 sessions, 28 participants. Parenthesized figures give sessions reaching verified firmware out of those run on that device, and are the same in both panels. Between-participant variance is 7.59 in Panel~A and 11.89 in Panel~B; in Panel~A, adding device raises it from 3.85, consistent with device and participant being partly confounded by the randomization. No device coefficient is interpreted: with 13 to 16 sessions per device, the intervals span three orders of magnitude or more, and all include~1.
\end{tablenotes}
\end{threeparttable}
\end{table}
\newpage
\section{Device Pathway Overview}
\label{app:ui_complx}

\begin{table}[H]
\centering
\caption{Pathways to a device-held credential, and password session outcomes. Pathway properties were mapped after data collection (Section~\ref{sec:prestudy}); outcome counts are the sessions reported in Section~\ref{sec:results}. ``None located'' records what the audit was able to identify from the device, its companion app, and its support materials, and does not establish that no such credential exists. Session counts differ by device because device assignment was randomized.}
\label{tab:password_change}
\begin{tabularx}{\textwidth}{l c Y c Y Y}
\toprule
\textbf{Device} & \textbf{Steps} & \textbf{Interfaces involved} & \textbf{Navigation depth} & \textbf{Feedback on change} & \textbf{Session outcomes ($n$)} \\
\midrule
Meross Smart Plug & --- & None located & --- & --- & Account 12, none 4 (16) \\
SwitchBot Hygrometer & --- & None located & --- & --- & Account 13, none 1 (14) \\
Arlo Chime 2 & --- & None located & --- & --- & Account 12, none 2 (14) \\
Amazon Echo Dot & --- & None located & --- & --- & Account 1, none 13 (14) \\
HP DeskJet 2721e & 6 & App + browser/EWS & 4 levels & Low & Device-held 1, account 3, none 9 (13) \\
TP-Link Tapo C210 & 5 & App only & 3 levels & None & Account 9, none 4 (13) \\
\bottomrule
\end{tabularx}
\begin{flushleft}
\footnotesize
Steps, interfaces, and navigation depth are counted from the app's home screen to the control that changes a credential held on the device. Feedback on change records whether the interface reported the resulting state. Em dashes indicate that no such pathway was found, and the property is therefore undefined. Session outcomes: \emph{device-held} = a session reached the control that changes a credential stored on the device; \emph{account} = a session reached a credential setting on an online account; \emph{none} = no password setting was reached. 
\end{flushleft}
\end{table}

\begin{table}[ht]
\centering
\caption{Pathways to firmware status, and update session outcomes. Pathway properties were mapped after data collection; outcome counts are the sessions reported in Section~\ref{sec:results}. Session counts differ by device because device assignment was randomized.}
\label{tab:firmware_update}
\begin{tabularx}{\textwidth}{l c Y c Y Y}
\toprule
\textbf{Device} & \textbf{Steps\(^1\)} & \textbf{Interfaces involved} & \textbf{Navigation depth} & \textbf{Terminology matches advice} & \textbf{Session outcomes ($n$)} \\
\midrule
Amazon Echo Dot & 2--3 & Voice only & 1 level & No\(^2\) & Firmware 5, app 2, none 7 (14) \\
Meross Smart Plug & 3--4 & App only & 2 levels & Yes & Firmware 9, app 7, none 0 (16) \\
SwitchBot Hygrometer & 4--5 & App only & 2 levels & Yes & Firmware 4, app 3, none 7 (14) \\
Arlo Chime 2 & 5--6 & App only & 3 levels & Yes & Firmware 6, app 2, none 6 (14) \\
HP DeskJet 2721e & 4--5 & App + browser/EWS & 4 levels & No & Firmware 3, app 4, none 6 (13) \\
TP-Link Tapo C210 & 5--6 & App only & 4 levels & Yes & Firmware 11, app 1, none 1 (13) \\
\bottomrule
\end{tabularx}
\begin{flushleft}
\footnotesize
Steps, interfaces, and navigation depth are counted from the app's home screen (or, for the Echo Dot, from the first utterance) to the point at which firmware status is reported. Terminology matches advice records, whether the interface labels this function in the vocabulary the advice uses. Devices are ordered by navigation depth. Session outcomes: \emph{firmware} = a session reached a point at which the device's firmware status was reported; \emph{app} = a session reached a companion-application update; \emph{none} = neither was reached.\\
\(^1\) The number of steps varies by one depending on whether an update is available; if one is detected, an additional confirmation step is required.\\
\(^2\) Reaching the status report requires a specific phrasing (for example, ``Check for software updates'' rather than ``Update''); the pathway is short but is not browsable, so the wording cannot be discovered by inspection.
\end{flushleft}
\end{table}
\clearpage
\section{Session Outcomes by Participant Characteristic}
\label{appendix:demographics}

\begin{table}[!htbp]
\centering
\captionsetup{
    width=\linewidth,
    justification=justified,
    singlelinecheck=false
}
\caption{Distribution of session outcomes across participant characteristics, reported for completeness rather than as subgroup comparisons. Each participant contributed three password sessions and three update sessions, so session denominators do not represent independent observations. The subgroups are strongly imbalanced, with 22 of 28 participants in the highest education band and one age band containing fewer than three participants; device assignment was randomized, so groups did not encounter the same devices. Percentages are omitted because the smallest cells are based on one or two participants. Outcome categories are those defined in \Cref{sec:coding} and reported by device in \Cref{tab:results_overview}. Education follows the three-way Dutch classification described in \Cref{sec:participant_recruitment}; prior IoT experience denotes self-reported use of at least one IoT device at home, at work, or through friends or relatives.}
\label{tab:outcomes_by_characteristic}

\begin{threeparttable}

\footnotesize

{\raggedright Panel A. Password advice (84 sessions)\par}\smallskip

\begin{tabular*}{\linewidth}{@{\extracolsep{\fill}}llcccc@{}}
\toprule
 & & & \multicolumn{3}{c}{Sessions reaching} \\
\cmidrule(lr){4-6}
Characteristic & Group & Participants & No password & Account-level & Device-level \\
 & & ($n$) & setting & setting & setting \\
\midrule
Education  & Low    & 1  & 2  & 1  & 0 \\
           & Middle & 5  & 4  & 11 & 0 \\
           & High   & 22 & 27 & 38 & 1 \\
\midrule
Age        & $<$30  & 3  & 3  & 6  & 0 \\
           & 30--39 & 2  & 3  & 3  & 0 \\
           & 40--49 & 6  & 2  & 15 & 1 \\
           & 50--59 & 6  & 5  & 13 & 0 \\
           & 60+    & 11 & 20 & 13 & 0 \\
\midrule
Prior IoT  & Yes    & 20 & 23 & 36 & 1 \\
experience & No     & 8  & 10 & 14 & 0 \\
\midrule
All        &        & 28 & 33 & 50 & 1 \\
\bottomrule
\end{tabular*}

\medskip

{\raggedright Panel B. Update advice (84 sessions)\par}\smallskip

\begin{tabular*}{\linewidth}{@{\extracolsep{\fill}}llcccc@{}}
\toprule
 & & & \multicolumn{3}{c}{Sessions reaching} \\
\cmidrule(lr){4-6}
Characteristic & Group & Participants & No update & Companion-app & Verified firmware \\
 & & ($n$) & reached & update & update \\
\midrule
Education  & Low    & 1  & 2  & 0  & 1  \\
           & Middle & 5  & 4  & 6  & 5  \\
           & High   & 22 & 21 & 13 & 32 \\
\midrule
Age        & $<$30  & 3  & 1  & 0  & 8  \\
           & 30--39 & 2  & 0  & 3  & 3  \\
           & 40--49 & 6  & 6  & 3  & 9  \\
           & 50--59 & 6  & 5  & 8  & 5  \\
           & 60+    & 11 & 15 & 5  & 13 \\
\midrule
Prior IoT  & Yes    & 20 & 20 & 10 & 30 \\
experience & No     & 8  & 7  & 9  & 8  \\
\midrule
All        &        & 28 & 27 & 19 & 38 \\
\bottomrule
\end{tabular*}

\begin{tablenotes}[flushleft]\footnotesize
\item Each group contributes three sessions per participant per advice item, so the
session denominator for any row is three times its participant count.
\end{tablenotes}

\end{threeparttable}
\end{table}
\clearpage
\section{Synthesis of Advice Text}
\label{sec:traceability-matrix}

\begin{table*}[ht]
\centering

\caption{Source mapping and synthesis of password advice.}
\label{tab:password-advice}

\begingroup
\footnotesize
\begin{tabularx}{\textwidth}{p{3cm} p{4.5cm} X}
\toprule
\textbf{Actionable Element} & \textbf{Source National Campaigns (Consensus Rationale)} & \textbf{Consolidated Advice Text Provided to Participants} \\
\midrule
\textbf{Identifying the Target} & \textbf{CH:} ``Standard configuration'' \cite{Switzerland}. \par \textbf{DE:} ``Preset default passwords'' \cite{Germany}. \par \textbf{US:} ``Configured with default passwords to simplify setup'' \cite{US}. & ``A default password is a pre-configured password that comes with your device.'' \\
\addlinespace[6pt]
\textbf{Risk Attribution} & \textbf{AU:} ``Collected and posted online'' \cite{Australia}. \par \textbf{CH:} ``Can be easily hacked'' \cite{Switzerland}. \par \textbf{JP:} ``Data... publicly exposed'' \cite{Japan}. \par \textbf{US:} ``Easily found online... don't provide any protection'' \cite{US}. & ``These passwords are often simple, publicly known, and easily found online, making them a prime target for hackers.'' \\
\addlinespace[6pt]
\textbf{Security Consequences} & \textbf{DE:} ``Botnets... remote control'' \cite{Germany}. \par \textbf{JP:} ``Stepping stones for attacks'' \cite{Japan}. \par \textbf{US:} ``...often the only barrier between you and your personal information'' \cite{US}. & ``Leaving the default password unchanged can have serious consequences, including unauthorized access to your personal information, the installation of malware, and cyberattacks.'' \\
\addlinespace[6pt]
\textbf{Primary Instruction (Generic)} & \textbf{NL:} ``Change the default'' \cite{VeiligInternetten2}. \par \textbf{AU:} ``Important... to change it'' \cite{Australia}. \par \textbf{ES:} ``Remember to change'' \cite{Spain}. & \textbf{Action:} ``To protect yourself and your data, it is important to change the default password as soon as possible.'' \textit{(Note: No locational cues provided.)} \\
\bottomrule
\end{tabularx}
\endgroup

\end{table*}

\begin{table*}[ht]
\centering

\caption{Source mapping and synthesis of update advice.}
\label{tab:update-advice}

\begingroup
\footnotesize
\begin{tabularx}{\textwidth}{p{3cm} p{4.5cm} X}
\toprule
\textbf{Component of Study Advice} & \textbf{Source National Campaigns (Consensus)} & \textbf{Consolidated Advice Text Provided to Participants} \\
\midrule
\textbf{Rationale: Functionality \& Security} & 
\textbf{NL:} ``Importance for their functionality and security'' \cite{VeiligInternetten2}. \par
\textbf{ES:} ``Reaping benefits while minimizing risks'' \cite{Spain}. \par
\textbf{US:} ``Reap benefits if devices are secure and trusted'' \cite{US}. & 
``Manufacturers frequently release updates to improve the functionality and security of their smart devices.'' \\
\addlinespace[6pt]

\textbf{Threat Mitigation} & 
\textbf{JP:} ``Fix weaknesses (security holes)... called vulnerabilities'' \cite{Japan}. \par
\textbf{AU:} ``Updates to fix security vulnerabilities'' \cite{Australia}. \par
\textbf{US:} ``Patches to fix the problem... vulnerabilities in device's software'' \cite{US}. & 
``These updates help fix vulnerabilities, protect against cyber threats, and ensure that the device runs smoothly.'' \\
\addlinespace[6pt]

\textbf{Step 1: Access Device (App/Web)} & 
\textbf{NL:} ``Step 1: Open the app for your smart device'' \cite{VeiligInternetten2}. \par
\textbf{DE:} ``Via a dedicated app or the device's web interface'' \cite{Germany}. \par
\textbf{ES:} ``Access the device application through your mobile device'' \cite{Spain}. & 
``\textbf{Access your device:} Open the device app or the configuration menu/web interface.'' \\
\addlinespace[6pt]

\textbf{Step 2: Navigation (Settings)} & 
\textbf{NL:} ``Step 2: Go to 'Settings' (Instellingen)'' \cite{VeiligInternetten2}. \par
\textbf{US:} ``Examine the settings, particularly security settings'' \cite{US}. & 
``\textbf{Find Settings:} Locate the 'Settings' section.'' \\
\addlinespace[6pt]

\textbf{Step 3: Verification \& External Check} & 
\textbf{NL:} ``Step 3: Click on 'Update'... if no app, go to manufacturer's website'' \cite{VeiligInternetten2}. \par
\textbf{JP:} ``Regularly check the manufacturer's website'' \cite{Japan}. \par
\textbf{US:} ``Apply relevant patches as soon as possible'' \cite{US}. & 
``\textbf{Check the update status:} Select 'Update' or navigate to the updates section on the supplier's website to check if the latest update has been installed. If a new update is available, install it.'' \\

\bottomrule
\end{tabularx}
\endgroup

\end{table*}
\newpage
\section{Codes clustered into Themes and Subthemes}
\label{sec:themes}
\begin{longtable}{p{0.22\linewidth} p{0.28\linewidth} p{0.44\linewidth}}
\caption{Codes from the coding framework were clustered into analytic themes and subthemes. Each theme is centered on a central concept, presented in the first column; the subthemes capture participants’ actions and statements during the sessions. Themes were developed by interpretive synthesis rather than by counting frequencies, and the frequencies of themes and subthemes, together with their co-occurrence with session outcomes, are therefore not presented (Section~\ref{sec:qualitative}). Code labels are retained as recorded in the archived codebook.}
\label{tab:codebook}\\
\toprule
\textbf{Theme (central concept)} & \textbf{Subtheme} & \textbf{Codes (from coding framework)} \\
\midrule
\endfirsthead
\toprule
\textbf{Theme (central concept)} & \textbf{Subtheme} & \textbf{Codes (from coding framework)} \\
\midrule
\endhead
\bottomrule
\endlastfoot
\textbf{Interpreting What the Advice Means} \newline \emph{Which credential or update object the card refers to, and at which layer it sits, as inferred from the card and the interface} & Identifying the credential or update object the advice names & Password-Related Processes; Password Setting Identified as Target; Identifying Update Requirements \\
& Treating an account-level or companion-app action as the advice's target & Account-Level Password Workflow; Influence of Platform-Suggested Passwords; App Version Screen Treated as Firmware Status; Assumption of Security Feature Functionality; Familiarity with Platform; Familiarity with Device; Simulating Real-World Behavior \\
& Uncertainty about the target & Uncertainty About Existence of Security Feature; Unclear Distinction Between Device and App Updates; User Assumptions Based on Context \\
& Language mismatch / jargon & Challenges of Jargon and Labeling; Language Preference for Interface \\
\midrule
\textbf{Navigating Interfaces: Smooth \& Stuck Journeys} \newline \emph{Whether interface labels and structure led to a candidate setting} & Direct routes to a candidate setting & Quickly Locating Features; Efficient Pathways \\
& Unexpected or difficult-to-locate settings & Navigational Difficulties (Search Returning No Setting, Looping in Support Pathways, Repeated Trial-and-Error, etc.); Challenges in Finding Specific Features \\
& Inconsistent or absent feedback & Unclear Interface Elements; Lack of Security Confirmation; Difficulty Locating Password/Update Options \\
& Trial-and-error navigation & Trial-and-Error in Troubleshooting; Restarting App \\
& Device contrasts & Comparing Devices Based on Security Usability \\
\midrule
\textbf{The Question No Source Could Answer} \newline \emph{Support materials could describe how to perform an action but not whether it existed on the device at hand; every channel presupposed the target had been identified} & Support that presupposed the target & Help and Support Section; Virtual Assistant; User Guide; Unhelpful Instructions \\
& Adjudicating between third parties & Google; Manufacturer Website; E-mail Support \\
& The contrasting case: a channel that answered the question & Helpful Instructions; Recognizing Previously Seen Information \\
\midrule
\textbf{Searching Without a Stopping Rule} \newline \emph{Nothing signalled when a task was finished or could not be finished; the search ran until abandoned, ended without confirmation, and its absence of an endpoint was read as a judgment about oneself} & Continuing without an endpoint & Complexity and Effort Perception (Mental Workload, Effort Evaluation, Perceived Time Investment); Persistent Search for Information \\
& Stopping by abandonment & User Experience: Fatigue and Frustration (Fatigue; Abandonment) \\
& Absorbing the absence of an endpoint as a judgment about oneself & User Experience: Fatigue and Frustration (Emotional Reactions) \\
& Ending without confirmation & Mixed Feelings About Having Applied the Advice; Unclear Outcome \\
& Boundary case: sessions that ended at an explicit status & Smooth Interaction; Confidence with Little Effort; Satisfaction with Outcome \\
& Prior stance on the value of securing a small device (interview context) & Skepticism About Security Measure Effectiveness; Perceived Value of Additional Authentication \\

\end{longtable}

\end{document}